\documentclass[11pt]{article}

\usepackage[final]{acl}
\usepackage{multirow}
\usepackage{stfloats}
\usepackage[table]{xcolor}
\definecolor{darkgreen}{RGB}{0,100,0}
\usepackage[normalem]{ulem}
\usepackage{lmodern}
\usepackage{orcidlink}
\usepackage{times}
\usepackage{latexsym}
\usepackage{booktabs} 
\usepackage{placeins}
\usepackage[most]{tcolorbox}
\usepackage{xcolor}
\newtcolorbox{promptbox}[1]{
  colback=gray!3,
  colframe=gray!55,
  coltitle=black,
  fonttitle=\bfseries,
  title=#1,
  boxrule=0.5pt,
  arc=2mm,
  left=2mm,
  right=2mm,
  top=1mm,
  bottom=1mm,
  enhanced,
  breakable
}
\usepackage{xurl}
\usepackage{hyperref}
\usepackage{tabularx}
\usepackage[T1]{fontenc}
\usepackage[utf8]{inputenc}

\usepackage{microtype}

\usepackage{inconsolata}

\usepackage{graphicx}

\title{\textbf{STORM}: Stepwise Token Optimization with Reward-Guided Beam Search}

\author{
  \textbf{Arthur Satouf}\orcidlink{0009-0001-0198-3868}\textsuperscript{1,2,3,4},
  \textbf{Giulio D'Erasmo}\orcidlink{0009-0002-2287-051X}\textsuperscript{5},
  \textbf{Yuxuan Zong}\orcidlink{0009-0002-0376-1369}\textsuperscript{4},  \\
  \textbf{Habiboulaye Amadou-Boubacar}\orcidlink{0000-0001-8387-0068}\textsuperscript{3},
  \textbf{Pablo Piantanida}\orcidlink{0000-0002-8717-2117}\textsuperscript{1,2}, 
  \textbf{Benjamin Piwowarski}\orcidlink{0000-0001-6792-3262}\textsuperscript{4} 
  \\
  \textsuperscript{1} MILA – Quebec AI Institute \& ILLS, Canada \\
  \textsuperscript{2}Université Paris-Saclay \& CentraleSupélec \&  CNRS, France \\
  \textsuperscript{3}Air Liquide, France \\
  \textsuperscript{4}Sorbonne Université \& ISIR \& CNRS, France \\
  \textsuperscript{5}Sapienza, University of Rome, Italy \\
   \small{
  \texttt{arthur.satouf(at)gmail.com, g.derasmo@diag.uniroma1.it, zong@isir.upmc.fr,} }\\
  \small{
  \texttt{ habiboulaye.amadou-boubacar@airliquide.com, pablo.piantanida@cnrs.fr,  benjamin.piwowarski@cnrs.fr}
  }
}

\usepackage{amsmath, amssymb}
\usepackage{algorithm}
\usepackage{algpseudocode}

\begin{document}

\maketitle

\begin{abstract}

Modern retrieval increasingly relies on dense and learned-sparse neural
  models that are effective but require encoding the entire corpus into a
  specialized index, rebuilt whenever the model changes.
  Lexical retrievers like BM25 stay efficient and transparent on a
  standard inverted index that need not change as models evolve, but
  suffer from vocabulary mismatch. LLM query rewriting can help, yet
  prompted rewriters emit well-formed but retrieval-ineffective---or
  harmful---terms, and training against a retrieval reward gives only
  delayed, sequence-level supervision that obscures which terms helped.
  We introduce \textbf{STORM} (\textbf{S}tepwise \textbf{T}oken
  \textbf{O}ptimization with \textbf{R}eward-guided bea\textbf{M} search),
  a self-supervised with retrieval feedback framework for lexical query expansion. STORM trains
  the rewriter through generation guided by retrieval metrics: at each
  step, candidate expansions are scored against the BM25 index
  and low-reward continuations pruned, turning the retrieval reward
  into a token-level signal that concentrates exploration on
  retrieval-effective vocabulary.
  Across TREC DL and BEIR, STORM lets 0.6B--8B backbones match or surpass
  competitive LLM rewriters while retrieving as fast as plain BM25; at 8B
  it rivals far larger proprietary rewriters. It further transfers
  zero-shot to 18 languages (MIRACL), beating dedicated multilingual dense
  retrievers on average, making STORM a competitive, infrastructure-light
  alternative to dense neural retrieval\footnote{EMNLP 2026 - Budapest - Main Conference}.

\end{abstract}

\section{Introduction}

Information Retrieval (IR) retrieves, from a large collection, the documents relevant to a user's information need, and is a key component of search engines and of Retrieval-Augmented Generation (RAG)~\citep{lewis2020retrieval}.
% Keyword-based retrieval remains one of the most widely deployed paradigms for large-scale search, typically as the first-stage retriever.
Despite the strong performance of neural retrievers~\cite{nogueira2019passage, karpukhin2020dense, formal2021splade}, lexical methods such as BM25~\cite{Robertson1994OkapiAT} retain important practical advantages, notably highly efficient retrieval through optimized index structures such as Block-Max WAND~\cite{grand_maxscore_2020}.
Their effectiveness, however, is constrained by vocabulary mismatch~\cite{10.1016/j.ipm.2019.05.009}, where relevant documents use terms absent from the user’s query---motivating lexical expansion and query reformulation prior to retrieval.
% The following part seems to be overlapping part. 
Prior work tackles this gap by rephrasing the input or injecting lexical terms, mostly in two regimes: Pseudo-Relevance Feedback (PRF) methods that pick expansion terms from top-ranked documents via hand-crafted scoring rules (e.g., RM3~\cite{lavrenko2001relevance}),
and LLM-based rewriters that rewrite the query~\cite{jagerman2023queryexpansionpromptinglarge}, generating keywords~\cite{wang2023generativequeryreformulationeffective} or a document snippet~\cite{wang2023query2docqueryexpansionlarge} via prompting~\cite{gao2023precise, shen2024large, ma2023query} or supervised finetuning~\cite{mao2024rafe}---without consulting the retriever during generation.
More recent work~\cite{wang2025maferw, chan2024rq, satouf-etal-2026-quester} uses reinforcement learning to train an LLM rewriter end-to-end against a retrieval reward, dynamically adapting queries to retriever feedback.

However, in RL-based rewriting the reward is observed only \emph{after} the full sequence is generated, making it hard to tell which lexical terms help retrieval and which do not.
This creates an exploration bottleneck: the space of keyword sequences is large and high-reward rewrites are rare---most completions mirror the original query, add generic terms, or drift toward semantically related yet retrieval-ineffective vocabulary. Sequence-level on-policy methods thus receive delayed, weak supervision, often converging slowly or collapsing onto a few expansions.

We address this with reward-guided beam search. Our key insight is that beam search can serve not only as an inference-time decoder but as a \emph{training-time} exploration mechanism. We introduce \textbf{STORM} (\textbf{S}tepwise \textbf{T}oken \textbf{O}ptimization with \textbf{R}eward-guided bea\textbf{M} search): at each generation step, candidate keyword sequences are scored by a retriever, and only those achieving high relevance are retained for further expansion. This concentrates exploration on retrieval-effective vocabulary, unlike plain beam search, which collapses onto a few high-likelihood but retrieval-ineffective completions.% at each expansion step, every partial keyword sequence is decoded, submitted to a BM25 index, and scored with a composite retrieval reward (Section~\ref{sec:reward}). Branches whose partial vocabularies are unlikely to retrieve relevant documents are pruned, while promising branches are expanded further. The resulting samples concentrate probability mass on high-reward regions of the output space, in contrast to standard policy sampling, which wastes its budget on low-reward regions, and plain beam search, which collapses onto a few high-likelihood completions.

We train the policy of the LLM using the reward-guided beam search as a structured proposal distribution, adapting the Generative Cooperative Networks (GCN) framework~\cite{lamprier2022generativecooperativenetworksnatural} by replacing its learned discriminator with a predefined retrieval reward. The retrieval reward thus plays two roles: it guides exploration during decoding, and it reweights generated samples during the policy update. As a side benefit, STORM achieves lower inference latency than competing LLM-rewrite pipelines while improving retrieval effectiveness, since its expansions are short keyword strings rather than full pseudo-documents. \textbf{STORM converts retrieval rewards into token-level lexical exploration through reward-guided beam search.}

\medskip
\noindent Our contributions are:
\begin{itemize}
    \item A self-supervised with retrieval feedback training method for lexical query rewriting requiring no human-written rewrites: supervision comes entirely from the retriever's scoring of generated expansions.
    \item \emph{Reward-guided stochastic beam search}, a step-wise exploration mechanism that turns retrieval feedback into a token-level signal, helping the model discover retrieval-effective vocabulary during generation.
    \item Empirical evidence across in-domain (TREC DL collections) and out-of-domain (BEIR) benchmarks showing consistent gains and lower inference cost than LLM-rewrite + BM25 baselines.\footnote{Code: \url{https://github.com/arthur-75/storm}. 
Trained models: \url{https://huggingface.co/collections/Arthur-75/storm}.}
    \item Zero-shot cross-lingual transfer to 18 languages (MIRACL), despite training only on English MS-MARCO data, outperforming dedicated multilingual dense retrievers on average.
\end{itemize}
 
\medskip
\noindent The remainder of the paper is organized as follows: Section~\ref{sec:method} formalizes the task and the method, Section~\ref{sec:experiments} presents the experimental setup and results. We conclude in Section~\ref{sec:conclusion}.

% This section is organized as follows. Section~3.1 formalizes the task and the token-to-term mapping $T$. Section~3.2 defines the reward function. Section~3.3 derives the training objective and shows how the GCN importance-sampling framework yields a stable, self-normalized gradient estimator in our setting. Section~3.4 specifies the proposal distribution used at training time.

\section{Related Work}

\paragraph{Query Rewriting in IR.}

Query rewriting is a long-standing technique in information retrieval designed to close the vocabulary gap between a user’s expressed query and the terms used in relevant documents~\cite{lavrenko2001relevance}. 
% Classical approaches—pseudo-relevance feedback, thesaurus expansion, and term weighting—are effective but brittle when queries are short, under-specified, or use colloquial phrasing.  Modern retrieval pipelines increasingly delegate this step to neural models~\cite{vakulenko-etal-2020-wrong, yates-etal-2021-pretrained}leveraging the world-knowledge encoded in pre-trained language models to produce richer, semantically aligned queries expansions.
Early approaches, such as pseudo-relevance feedback (PRF)~\cite{lavrenko2001relevance, abdul-jaleel2004umass}, perform query rewriting by leveraging heuristic or statistical analyses of documents retrieved using the original query. 
Because the initial retrieval set may contain false positives, these methods are particularly susceptible to query drift~\cite{mitra1998improving}, where the reformulated query gradually deviates from the user’s original intent. As a result, reformulation precision may deteriorate, ultimately reducing overall retrieval effectiveness.
More recent work proposes to rewrite the query with LLMs, taking advantage of their strong semantic understanding and broad knowledge. One direct approach simply prompts the LLM to rewrite or expand the query into various forms, including the pseudo-query~\cite{alaofi2023can, mao2024rafe}, the keywords~\cite{jagerman2023queryexpansionpromptinglarge,ma2023query}, or a longer answer-bearing passage~\cite{gao2023precise,wang2023query2docqueryexpansionlarge,shen2024large,zhang2024exploring}. HyDE~\cite{gao2023precise} generates a single hypothetical pseudo-document from a zero-shot prompt and uses it as the expanded query, relying on the LLM's parametric knowledge to bridge the vocabulary gap. MuGI~\cite{zhang2024exploring} extends this idea by generating multiple pseudo-references and integrating them with the original query through a dynamic length-based repetition ratio, enriching the lexical coverage of the expansion. W2P~\cite{choi-etal-2025-word2passage} further refines the integration by generating pseudo-references at three granularities---word, sentence, and passage---then re-weighting individual words using query-type-specific significance scores tuned via grid search. Despite their increasing sophistication, these methods all share the same fundamental limitation: the expansion process is decoupled from the retriever, and generating long pseudo-references, sometimes multiple per query, incurs substantial inference latency. The LLM never observes whether its terms actually improve ranking, and expansions are calibrated by prompt design or hyperparameter search rather than retrieval feedback.

A separate line of work~\cite{wang2025maferw,satouf-etal-2026-quester} uses reinforcement learning to optimize LLMs against sequence-level retrieval feedback, aiming to generate rewrites that better align with the underlying information need. However, this sequence-level framing introduces two limitations. First, the retrieval reward is observed only once the full rewrite has been generated: QUESTER~\cite{satouf-etal-2026-quester} assigns a single SoftNDCG score---a smooth, differentiable approximation of nDCG---to each complete keyword sequence, so the model never learns which individual lexical choices were responsible for the gain. Second, exploration is left entirely to stochastic sampling at high temperature, with no structural bias toward retrieval-effective vocabulary. STORM addresses both issues by turning the retrieval reward into a token-level signal that prunes unpromising branches at each generation step, rather than scoring complete sequences after the fact.

%We frame query rewriting as a keyword specification query rewrite sequence-to-sequence generation problem simillaer to ~\cite{satouf-etal-2026-quester} to  optimize against a retrieval reward. Given a user query $x$, an autoregressive language model $p_\theta$ produces a rewritten query $\hat{y}$, which is then scored by issuing it to a keyword search engine, here BM25~\cite{Robertson1994OkapiAT} and evaluating the retrieved documents against relevance judgments. Because the reward is computed on the \textit{terms} fed to BM25---not on the LLM token sequence directly---the training objective involves a non-differentiable mapping between two vocabularies. We address this by adapting the gradient estimator and sampling scheme of Generative Cooperative Networks (GCN; Lamprier et al., 2022), replacing GCN's learned discriminator with a fixed retrieval-based reward squashed into a probability-like signal.

\section{Methodology}
\label{sec:method}

\begin{figure}[h]
    \centering
    \includegraphics[width=1\linewidth, trim=.61cm 1.7cm 6.1cm .8cm, clip]{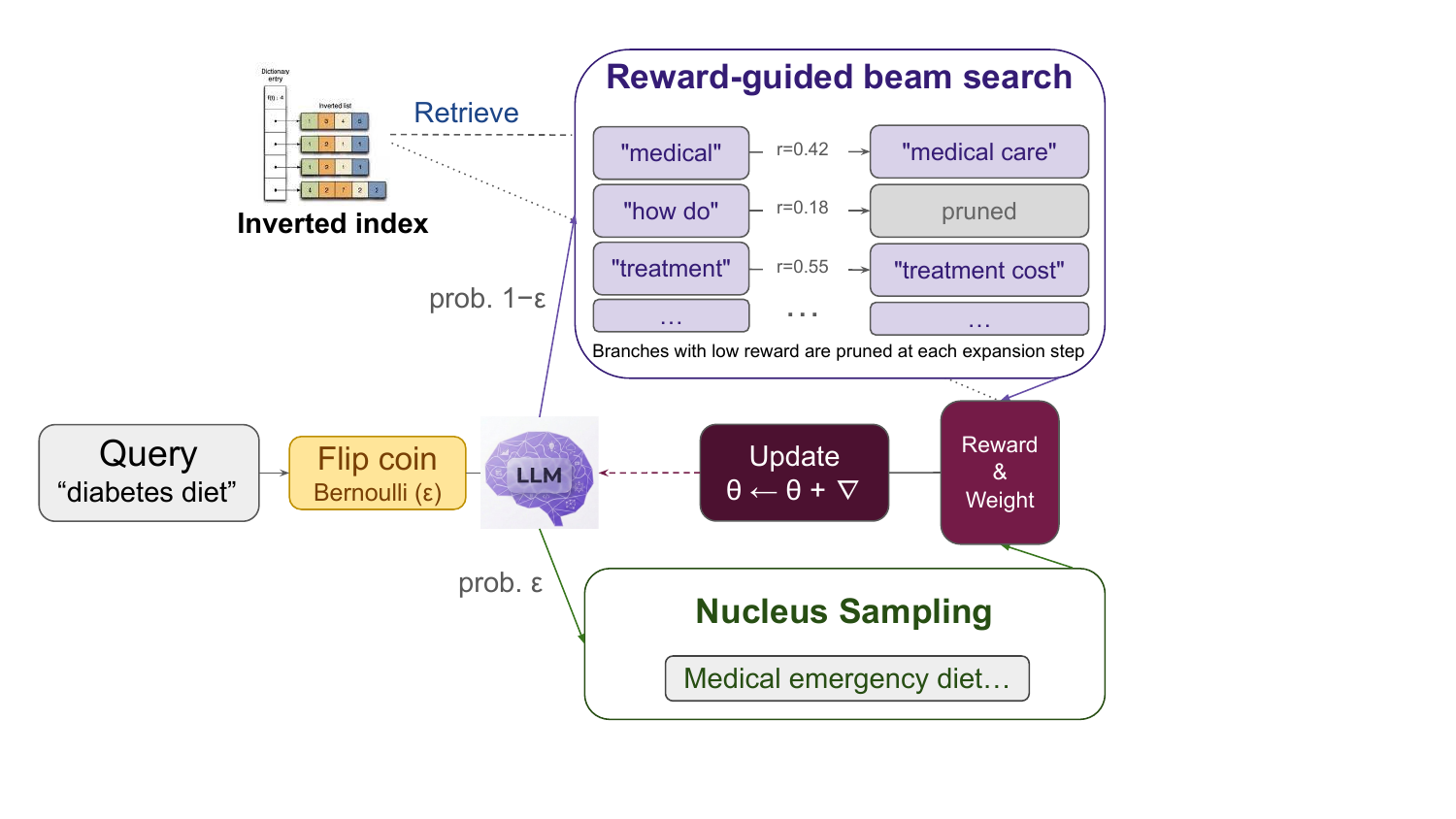}
       
    \caption{Overview of the STORM training loop. A Bernoulli($\varepsilon$) coin flip routes each query to either reward-guided beam search, where partial expansions are scored against the inverted index and low-reward branches are pruned, or nucleus sampling (classic generation). The selected sequence is then scored and importance-weighted to update the policy $\theta$.}
%, focusing exploration on retrieval-effective vocabulary
    \label{fig:storm}
\end{figure}
\newcommand{\query}{q}
\newcommand{\SE}{se}
\newcommand{\policy}{p_{\theta}}

\subsection{Task Formulation}

Let $\mathcal{V}_{\text{LLM}}$ denote the LLM token vocabulary and $\mathcal{V}_{\text{BM25}}$ the term vocabulary used by the retrieval system (typically a stemmed, lowercased word vocabulary). The LLM defines a conditional distribution over token sequences,

\begin{equation}
    p_\theta(y \mid x) = \prod_{j=1}^{|y|} p_\theta(y_j \mid x, y_{<j}),
\end{equation}

where $x$ is the input query and $y = (t_1, \ldots, t_{|y|})$ with $t_j \in \mathcal{V}_{\text{LLM}}$.

We define a deterministic mapping $T : \mathcal{V}_{\text{LLM}}^* \to \mathcal{V}_{\text{BM25}}^*$ that converts a token sequence into a sequence of retrieval terms. Formally, a contiguous block of tokens $(t_i, \ldots, t_{i+l})$ is mapped to a single term $w \in \mathcal{V}_{\text{BM25}}$ if and only if $t_{i+l} \in S$, where $S \subset \mathcal{V}_{\text{LLM}}$ is the set of delimiter tokens (whitespace, punctuation, end-of-sequence). Trailing tokens that have not yet encountered a delimiter are dropped by $T$, ensuring that only fully-formed terms are passed to BM25. For a partial generation $\hat{y}^{(k)} = (t_1, \ldots, t_k)$, we therefore have $T(\hat{y}^{(k)}) = (w_1, \ldots, w_m)$ with $m \leq k$.

Let $\hat{y}$ denote the rewritten query issued to BM25, which returns a ranked list of documents $\mathcal{D}_{\hat{y}}$. The ranking is then scored against the gold relevance judgments associated with the original query $x$.

\subsection{Training Objective}

Cooperative training frameworks such as GCN~\cite{lamprier2022generativecooperativenetworksnatural} train a policy $p_\theta$ with a joint learned discriminator $D_\phi : \mathcal{X} \times \mathcal{Y} \to [0, 1]$ that distinguishes high-quality from low-quality outputs. Training proceeds by minimizing $\text{KL}(q_\theta \,\|\, p_\theta)$, where the target distribution is

\begin{equation}\label{eq:q_def}
    q_\theta(y \mid x) \propto p_\theta(y \mid x) \, D_\phi(x, y). 
\end{equation}

We adopt this objective---which keeps the policy $p_\theta$ inside the target $q_\theta$---for two reasons. First, on-policy RL alternatives such as PPO or GRPO instead minimize the reverse KL $\text{KL}(p_\theta \,\|\, q)$ with high-variance score-function (REINFORCE) gradients. Second, an energy-based target that drops the policy, $q \propto \exp(D_\phi)$, tends to destabilize training once the discriminator saturates. \citet{lamprier2022generativecooperativenetworksnatural} show that retaining the policy inside the target avoids both problems and yields a convergent update.

The discriminator in Eq.~\ref{eq:q_def} serves only to assign each generated output a quality score in $[0, 1]$ that reweights samples toward high-quality regions, and the convergence argument above relies on keeping the policy inside the target, not on how this score is produced. In query rewriting we can measure output quality directly: a rewrite is good to the extent that it retrieves relevant documents---a signal grounded in relevance judgments rather than estimated by a learned model. We therefore replace $D_\phi$ with a predefined retrieval-based reward $R(\hat{y}, x)$, passed through a temperature-scaled sigmoid to map it to the same bounded range $[0, 1]$ that $D_\phi$ would occupy:

\begin{equation*}
% \tilde{R}_i = \sigma\left(k(R_i - \mu)\right) = \frac{1}{1 + \exp(-k(R_i - \mu))},
\tilde{R} = \sigma\left(k(R - \mu)\right) = \frac{1}{1 + \exp(-k(R - \mu))},
\end{equation*}

\noindent where $k$ and $\mu$ are fixed scalar hyperparameters, set empirically and not updated during training. The offset $\mu$ centers the reward distribution and the gain $k$ controls the sharpness of the sigmoid.

Substituting $\tilde{R}$ for $D_\phi$ in Eq.~\ref{eq:q_def} yields the target $q_\theta(y \mid x) \propto p_\theta(y \mid x)\tilde{R}(y, x)$.

Because direct sampling from $q_\theta$ is intractable, we estimate $\nabla_\theta \text{KL}(q_\theta \,\|\, p_\theta)$ via importance sampling from a tractable proposal $\hat{q}$. %Sampling directly from $p_\theta$ produces noisy, often degenerate outputs early in training, while pure beam search lacks the full support that importance sampling requires. We therefore use a mixture proposal
Following the original implementation, we use a mixture proposal

\begin{equation}\label{eq:proposal}
\hat{q}(\hat{y} \mid x) = \varepsilon \, p_\theta(\hat{y} \mid x) + (1 - \varepsilon) \, p^{\text{rgb}}(\hat{y} \mid x),
\end{equation}

\noindent where $p^{\text{rgb}}(\cdot \mid x)$ places all of its mass on $y^\star = \arg\max_{y \in B_K(x)} R(y, x)$, the highest-reward sequence in the beam set $B_K(x)$ returned by reward-guided beam search.

%\begin{equation*}
%p^{\text{rgb}}(y \mid x) = 
%\begin{cases}
%1/K, & y \in B_K(x), \\
%0, & \text{otherwise}.
%\end{cases}
%\end{equation*}

Sampling from $\hat{q}$ proceeds as follows: with probability $\varepsilon$, draw a sequence by nucleus sampling from $p_\theta$; with probability $1 - \varepsilon$, run beam search of width $K$ and select the highest-reward beam from $B_K(x)$ (we ablate this choice in Section~\ref{sec:ablation}). Figure~\ref{fig:storm} illustrates this training loop, and the full update rule is given in Algorithm~\ref{alg:training}.

\begin{algorithm}[H]
\caption{Training for Query Rewriting}
\label{alg:training}
\begin{algorithmic}[1]
\Require Policy $p_\theta$, reward function $R$, proposal $\hat{q}$, training set $\Gamma$, batch size $m$, learning rate $\eta$
\For{$t = 1, \ldots, T$}
    \State Sample $\{x^i\}_{i=1}^m$ from $\Gamma$
    \State $\forall i \in [\![1\,;m]\!]$: sample $\hat{y}^i \sim \hat{q}(\hat{y}^i \mid x^i)$
    \State
    \[
        \theta \;\leftarrow\; \theta
        + \eta\,\frac{1}{\sum_{j=1}^m w^j}
        \sum_{i=1}^m w^i \, \nabla_\theta \log p_\theta(\hat{y}^i \mid x^i)
    \]
    \State with
    \[
        w^i = \frac{p_\theta(\hat{y}^i \mid x^i)\, \tilde{R}(\hat{y}^i, x^i)}{\hat{q}(\hat{y}^i \mid x^i)}
    \]
\EndFor
\end{algorithmic}
\end{algorithm}

\paragraph{Fixed-reward interpretation.}
Unlike GCN, our quality function is fixed rather than jointly learned.
This changes the original GCN setting, but yields a simple interpretation
under idealized exact updates. Let $p_0$ be the initial policy and let
$\tilde R(y,x)>0$ be fixed. Repeated application of the target update gives

\[
p_t(y\mid x)
=
\frac{p_0(y\mid x)\tilde R(y,x)^t}
{\sum_{y'}p_0(y'\mid x)\tilde R(y',x)^t}.
\]

For bounded-length output spaces and positive initial support, as
$t\rightarrow\infty$ the probability mass therefore concentrates on the
maximizers of $\tilde R(\cdot,x)$. This argument concerns the idealized
fixed-reward target update; practical training additionally uses finite-sample
importance weighting and the likelihood component of our generation reward.
\subsection{Reward Function \label{sec:reward}}

The reward $R(\hat{y}^{(k)})$ combines retrieval effectiveness with a penalty for uninformative vocabulary, leading to more efficient retrieval:

\begin{align*}
R(\hat{y}^{(k)}) &= \text{nDCG}(T(\hat{y}^{(k)})) \\
&- \lambda_{\text{DF}} \sum_{w \in T(\hat{y}^{(k)})} \text{DF}(w) \\
&+ \lambda_L \log p_\theta(\hat{y}^{(k)} \mid x).
\end{align*}

The first term, nDCG, measures retrieval quality. Following QUESTER~\cite{satouf-etal-2026-quester}, we replace the sparse click-based MS-MARCO labels with the cross-encoder pseudo-labels from the OpenSearch Project MS-MARCO Hard Negatives LLM Scores\footnote{\url{https://huggingface.co/datasets/opensearch-project/msmarco-hard-negatives-llm-scores}} release~\cite{geng2024competitivesearchrelevanceinferencefree}. The nDCG@100 component of our reward is computed against these denser labels, effectively distilling the cross-encoder's relevance preferences into our lexical rewriter through the policy gradient.

The second term penalizes overly common terms, where $\text{DF}(w)$ is the normalized document frequency of term $w$ in the corpus; this discourages the model from emitting high-frequency terms that retrieve large, undiscriminating result sets. 

The third term, log-likelihood of the generated sequence under
the current policy, plays two roles: it imposes length regularization,
since $\log p_\theta$ scales with sequence length, and it provides a
useful prior for selecting the first generated token of each word, where the
retrieval reward is otherwise uninformative.

\section{Experiments, Results and Analysis}
\label{sec:experiments}

\begin{table*}[t]
\centering
\scriptsize
\setlength{\tabcolsep}{3pt}
\renewcommand{\arraystretch}{1.15}

\resizebox{\textwidth}{!}{%
\begin{tabular}{llccc|cccccccccccc|c}
\toprule
\multirow{2}{*}{LLM} & \multirow{2}{*}{Model}
& \multicolumn{3}{c|}{In-domain (MS-MARCO)}
& \multicolumn{12}{c|}{Out-of-domain (BEIR) }
& \multirow{2}{*}{\begin{tabular}{@{}c@{}}Avg.\\OoD\end{tabular}} \\
%& \multirow{2}{*}{Avg. OoD} \\
\cmidrule(lr){3-5} \cmidrule(lr){6-17}
& & DL-19 & DL-20
& Dev & NFC & SCID & SciF & Touché & FiQA
& Covid & Signal &News & RB04
& HotpotQA & NQ & DBP
& \\
\midrule
\multicolumn{2}{l}{Num. queries}
& 43 & 54
& 6980 & 323 & 1000 & 300 & 49 & 648
& 50 & 97 & 57 & 249
& 7405 & 3452 & 400
& \\

\midrule
\multirow{3}{*}{No}
& BM25
& 50.6 & 48.0 & 18.4 & 32.2 & 14.9 & 67.9 & 44.2 & 23.6
& 59.5 & 33.1 & 39.5 & 40.8 & 63.3 & 30.6 & 31.8 & 40.1 \\
& RM3
& 52.2 & 49.0 & 15.7 & 33.1 & 14.7 & 64.6 & 35.0 & 19.2
& 59.3 & 31.5 & 42.6 & 44.3 & 51.3 & 23.1 & 30.8 & 37.5 \\
& SPLADE-v2
& \textcolor{darkgreen}{\textbf{73.0}} &\textcolor{darkgreen}{\textbf{ 72.0 }}&\textcolor{darkgreen}{\textbf{ 38.3}} & 34.7 & \textcolor{darkgreen}{\textbf{15.9}} & 70.4 & 24.7 & \textcolor{darkgreen}{\textbf{34.7}}
& 72.7 & 30.1 & 41.5 & 46.8 & \textcolor{darkgreen}{\textbf{68.7}} & \textcolor{darkgreen}{\textbf{53.8 }}& \textcolor{darkgreen}{\textbf{43.7 }}& 44.8 \\

\midrule
\multirow{5}{*}{8B}

%& STORM-LoRA& 67.9 & 66.4 & 22.8 & 36.7 & 15.8 & 71.9 & 46.1 & 28.6 & 77.1 & 34.4 & 49.6 & 53.9 & 66.0 & 46.8 & 39.7 & 47.2 \\
& HyDE
& 49.4 & 40.7 & 10.9 & 31.2 & 13.4 & 68.2 & 33.9 & 16.3
& 58.3 & 19.7 & 36.5 & 41.4 & 49.7 & 33.4 & 30.0 & 36.0 \\
& MuGI
& \textbf{69.1} & 61.7 & 20.7 & 35.4 & 15.3 & 71.7 & 46.3 & 24.5
& 69.0 & 34.8 & 46.3 & 48.4 & 65.7 & 44.8 & 39.8 & 45.2 \\
& W2P
& 66.0 & 59.9 & 20.0 & {36.6} & 15.4 & 70.8 & \textcolor{darkgreen}{\textbf{49.1}} & 26.3
& 74.4 &\textbf{ 35.0} & 47.7 & 50.4 & \textbf{67.0} & 45.2 & 39.3 & 46.4 \\
& QUESTER 
&- &- & -&36.3&	15.2&	71.3&	41.4	&26.3&	69.8	&30.5&	46.8	&53.1	&63.4	&42.3&	36.0	&44.4\\
& STORM\textsubscript{64}
& \uline{66.1 }& \uline{\textbf{67.3} }& \uline{\textbf{22.9} }& \uline{\textcolor{darkgreen}{\textbf{36.8 }}}& \uline{\textbf{15.8}} &\uline{\textbf{ 72.8} }& 44.4&\uline{ \textbf{28.4}}
& \uline{\textcolor{darkgreen}{\textbf{79.6}} }& 33.5 &\uline{ \textcolor{darkgreen}{\textbf{{50.2}}}} & \uline{\textcolor{darkgreen}{\textbf{56.0} }}& \uline{65.9 }& \uline{\textbf{46.3} }& \uline{\textbf{40.1 }}& \textcolor{darkgreen}{\textbf{47.5}} \\
\midrule
\multirow{6}{*}{4B}

%& STORM-LoRA & 62.7 & 64.3 & 22.7 & 36.5 & 15.6 & 71.6 & 47.6 & 27.7 & 74.5 & 31.2 & 46.3 & 53.0 & 63.4 & 43.3 & 38.7 & 45.8 \\
& HyDE
& 51.7 & 41.7 & 10.4 & 29.0 & 12.7 & 68.5 & 34.6 & 16.4
& 58.2 & 21.2 & 28.6 & 40.4 & 49.4 & 29.7 & 24.8 & 34.5 \\
& MuGI
& \textbf{66.4} & 57.8 & 20.5 & 35.7 & 15.0 & \textcolor{darkgreen}{\textbf{73.5}} & 45.6 & 24.5
& 67.2 & 34.4 & 43.8 & 47.3 & 64.2 & 41.9 & 38.2 & 44.3 \\
& W2P
& 62.7 & 58.8 & 19.6 & 36.1 & 15.6 & 71.5 & 47.6 & 24.9
&\textbf{ 75.3} & 34.6 & 47.0 & 49.8 & 65.6 & 43.3 & 38.5 & 45.8 \\
& QUESTER 
& 63.1 & 60.8 & \textbf{22.4} & 36.0 & 15.1 & 69.3 & \textbf{47.7} & \textbf{27.5}
& 73.6 & 34.8 & 45.3 & 51.7 & 64.3 & 43.0 & 38.8 & 45.6 \\
& STORM\textsubscript{64}&
\uline{63.7}	&\uline{\textbf{64.4}}	&22.3	&\uline{\textbf{36.4}}&	\uline{\textbf{15.5}}	&\uline{\textbf{72.2}}&	47.3	&\uline{\textbf{27.7}}&\uline{75.2}	&33.6	&\uline{\textbf{48.7}}	&\uline{\textbf{53.5}}	&\uline{64.5} &\uline{\textbf{44.7}}	&\uline{\textbf{39.6}}	&{46.6} \\
\midrule
\multirow{6}{*}{1.7B}

% & STORM-LoRA & 53.9 & 58.9 & 19.9 & 35.0 & 14.5 & 69.4 & 41.6 & 25.8 & 75.7 & 30.6 & 43.7 & 52.8 & -- & -- & 32.7 & 42.2 \\
& HyDE
& 42.7 & 34.9 & 8.7 & 27.6 & 11.8 & 65.7 & 30.6 & 15.6
& 54.9 & 17.8 & 31.7 & 39.6 & 43.8 & 24.4 & 24.2 & 32.3 \\
& MuGI
& 62.6 & 58.8 & 19.8 & 35.4 & 14.4 & 71.2 & 45.9 & 23.6
& 69.6 & 34.6 & 43.8 & 46.6 & 63.1 & 39.0 &\textbf{ 38.5} & 43.8\\  
& W2P
& 61.0 & 56.1 & 18.8 & 35.9 & 14.7 & 71.1 & \textbf{48.7 }& 23.5
& 74.7 & \textcolor{darkgreen}{\textbf{35.2}} & 45.2 & 48.2 & \textbf{63.2} & 38.6 & 37.3 & 44.7 \\ 
& QUESTER
&{53.2} &	{54.8} & -- & {34.9} &\textbf{ 15.4 }& 66.9 & 44.1 & 25.9
& 69.9 & 33.9 & 42.4 & 48.7 & 62.8 & 36.8 & 34.8 & 43.0 \\

& STORM\textsubscript{32}&
\uline{\textbf{63.3}}	&\uline{\textbf{ 60.2}} &	\uline{\textbf{21.7}}&	\uline{\textbf{36.2}}	&15.3	&\uline{\textbf{71.6}}&	47.3	&\uline{\textbf{27.9}}&	\uline{\textbf{75.2}}&	32.8&	\uline{\textbf{45.7}}	&\uline{\textbf{52.2}}&	63.0&	\uline{\textbf{40.2}}&	\uline{37.4	}&\textbf{45.4} \\
\midrule
\multirow{6}{*}{0.6B}

%& STORM-LoRA & 52.1 & 50.2 & -- & 32.8 & 14.1 & 67.0 & 42.8 & 23.5 & 66.7 & 31.9 & 41.6 & 44.0 & -- & -- & 33.0 & 39.8 \\
& HyDE
& 33.8 & 25.6 & 7.47 & 20.7 & 10.0 & 62.0 & 28.9 & 13.6
& 44.5 & 19.8 & 28.0 & 27.4 & 48.3 & 20.0 & 14.7 & 28.2 \\
& MuGI
& 54.1 & 54.8 & 19.8 & 33.8 & 14.3 & 68.0 & 43.8 & 23.0
& 67.2 & 33.7 & 40.5 & 41.2 &  \textbf{63.0 }& 35.9 & 34.7 & 41.6 \\
& W2P
& 55.9 & 51.1 & 19.0 & \textbf{ 34.3} & 14.5 & 69.4 & \textbf{ 46.5} & 22.7
& 68.4 & \textcolor{darkgreen}{\textbf{ 35.2 }}& \textbf{ 42.5} & 44.6 &{  62.4 }& 35.1 & 34.9 & 42.5 \\
& QUESTER
& 54.1	&55.3 & -- & 33.9 &\textbf{ 14.9}& 67.1 & 44.4 & \textbf{26.0}
& 69.3 & 32.6 & 41.5 & 48.1 & 62.3 & 36.2 & 34.0 & 42.5 \\

& STORM\textsubscript{32}
& \uline{\textbf{ 56.7}} & \uline{\textbf{ 57.2}} & \uline{\textbf{ 21.0}} & 32.5 & { 14.7 }& \uline{\textbf{ 70.2}} & 45.9 & \uline{{ 25.7}}
&\uline{ \textbf{ 72.5}} & 32.3 & 40.4 &\uline{\textbf{  48.5}} &  {62.7} &  \uline{\textbf{37.1} }& \uline{\textbf{ 34.2 }}& {\textbf{ 43.1}} \\

\bottomrule
\end{tabular}%
}
\caption{nDCG@10 on in-domain and out-of-domain; (MRR@10 on MS-MARCO Dev).
\textit{LLM}: Qwen3 backbone shared by all LLM rewriters;
\textit{Avg.\ OoD}: average over BEIR.
All LLM methods retrieve via a common Pyserini BM25. {\textcolor{darkgreen} {Green}}: best across all methods.
\textbf{Bold}: best per backbone. \underline{Underline}: statistically significant improvement over BM25 ($p<0.05$).
 }
 \label{tab:main_results}
\end{table*}
%% ─────────────────────────────────────────────────────────────────

%% ─────────────────────────────────────────────────────────────────
\subsection{Experimental Setup}
\label{sec:setup}
 
\paragraph{Training.} We instantiate STORM on four Qwen3 sizes~\cite{qwen3technicalreport} (0.6B, 1.7B, 4B, and 8B), each trained for one epoch on ${\approx}80$k MS-MARCO queries~\cite{craswell2021msmarco,bajaj2016msmarco} against a BM25 retrieval reward computed via Pyserini~\cite{lin2021pyserinieasytousepythontoolkit} (Section~\ref{sec:reward}). Full data, optimization, and hardware details are deferred to Appendix~\ref{sec:training_details}.

\paragraph{Evaluation benchmarks.}
We evaluate the models with three types of datasets.
\textbf{In-domain:} TREC DL'19 (43 queries)~\cite{craswell2020trec2019dl} and DL'20 (54
queries)~\cite{craswell2021trec2020dl}, and MS-MARCO dev (6\,980 queries).
\textbf{Out-of-domain (BEIR):}~\cite{thakur2021beirheterogenousbenchmarkzeroshot} 12 collections totalling 14\,030 queries:
NFCorpus (NFC), SCIDOCS (SCID), SciFact (SciF), Touché-2020 (Touché), FiQA, TREC-COVID (Covid), Signal-1M (Signal), TREC-News (News), Robust04 (RB04), HotpotQA, NQ, and DBpedia-Entity (DBP).
\textbf{Multilingual (MIRACL):}~\cite{zhang-etal-2023-miracl} Eighteen languages (Arabic, Bengali, English,
Spanish, Persian, Finnish, French, Hindi, Indonesian, Japanese, Korean,
Russian, Swahili, Telugu, Thai, Chinese, German, Yoruba), totalling
13\,495 queries.
 
\paragraph{Baselines.}
We compare against four families of methods:
\begin{itemize}
    \item 
    \textit{Lexical baselines (no LLM):} BM25, RM3~\cite{lavrenko2001relevance}---a PRF-based model,
    and SPLADE-v2~\cite{formal2021splade}.

    \item \textit{Prompt-based rewriters with Qwen3 (same backbone as STORM),}
    paired with BM25: HyDE~\cite{gao2023precise}, MuGI~\cite{zhang2024exploring},
    W2P~\cite{choi-etal-2025-word2passage}, and QUESTER~\cite{satouf-etal-2026-quester}.

    \item 
    \textit{Prompt-based rewriters with GPT-4.1,} covering a broad
    range of recent methods evaluated in the reproducibility study
    of~\cite{Bigdeli2026ARS}: GenQR~\cite{wang2023generativequeryreformulationeffective},
    GenQREnsemble~\cite{Dhole_2024}, QA-Expand,
    Q2K~\cite{jagerman2023queryexpansionpromptinglarge},
    Q2D (zero-shot)~\cite{wang2023query2docqueryexpansionlarge},
    LameR~\cite{shen-etal-2024-retrieval}, CSQE~\cite{lei-etal-2024-corpus}, and MuGI.

    \item \textit{Dense retrievers} for the multilingual track: mDPR~\cite{karpukhin-etal-2020-dense,zhang-etal-2021-mr},
    mContriever~\cite{izacard2022unsuperviseddenseinformationretrieval},
    mColBERT~\cite{bonifacio2022mmarcomultilingualversionms,khattab2020colbertefficienteffectivepassage}, and HyDE (Qwen3-8B).
\end{itemize}

\paragraph{Retrieval and metrics.}
All methods retrieve using BM25 (Pyserini/Lucene) with the same inverted index for fairness, unless the method uses its own index (e.g.\ SPLADE-v2).
We report nDCG@10 for TREC DL, BEIR, and MIRACL, and MRR@10
    for MS-MARCO dev.
Finally, query latency is measured as average time per query
    (mean $\pm$ std over five independent runs on the same hardware).
 
\paragraph{STORM inference.}
At inference time we decode with diverse beam search~\cite{vijayakumar2018diversebeamsearchdecoding}:
$K{=}6$ beams partitioned into $3$ groups with diversity penalty $\lambda_d{=}1$,
returning the $3$ highest-scoring sequences per query.
Decoding is greedy within each beam and capped at $32$, $64$, or $128$
generated tokens (denoted by STORM\textsubscript{32},
STORM\textsubscript{64}, and STORM\textsubscript{128}).
The three rewrite sequences are concatenated into a single keyword string and
issued to BM25 in one retrieval call.

\subsection{Main Results}

\label{sec:main}

\paragraph{In-domain effectiveness.}
Table~\ref{tab:main_results} (left) reports in-domain results. STORM consistently leads on DL'20 across all four model sizes, with gains over BM25 of +9.2/+19.3 points in nDCG@10. On MS-MARCO dev, STORM ranks first or within 0.1 MRR@10 of the best prompt-based rewriter at every scale. Notably, STORM maintains strong performance even at smaller model sizes, whereas other methods degrade significantly. SPLADE-v2 remains ahead on in-domain, benefiting from extensive MS-MARCO training with dense supervision.

\paragraph{Out-of-domain generalisation.}
Table~\ref{tab:main_results} (right) reports results on 12 BEIR collections. STORM achieves the highest mean nDCG@10 at every model size, with STORM-8B reaching 47.5 and leading or matching the best baseline on 9 of 12 datasets. The advantage is especially pronounced over HyDE, suggesting that reward-guided beam search generalises better than zero-shot pseudo-document generation and that the retrieval reward compensates for reduced model capacity. Even STORM-0.6B surpasses all 0.6B baselines on average.

 \begin{table*}[t]
\centering
{\fontsize{7.9pt}{10pt}\selectfont

\setlength{\tabcolsep}{3.5pt}
\begin{tabular}{l rrrrrrrrrrrrrrrrrr |r}
\toprule
\textbf{Method}
  & \textbf{ar} & \textbf{bn} & \textbf{en} & \textbf{es} & \textbf{fa}
  & \textbf{fi} & \textbf{fr} & \textbf{hi} & \textbf{id} & \textbf{ja}
  & \textbf{ko} & \textbf{ru} & \textbf{sw} & \textbf{te} & \textbf{th}
  & \textbf{zh} & \textbf{de} & \textbf{yo}
  & \textbf{Avg.} \\
{ \# queries}
  & { 2896} & { 411} & { 799} & { 648} & { 632}
  & { 1271} & { 343} & { 350} & { 960} & { 860}
  & { 213} & { 1252} & { 482} & { 828} & { 733}
  & { 393} & { 305} & { 119}
  & { } \\
\midrule
BM25
  & 48.1 & 50.8 & 35.1 & 31.9 & 33.3
  & 55.1 & 18.3 & 45.8 & 44.9 & 36.9
  & 41.9 & 33.4 & 38.3 & 49.4 & 48.4
  & 18.0 & 22.6 & 40.6
  & 38.5 \\
mDPR
  & 49.9 & 44.3 & 39.4 & \textbf{47.8} & \textbf{48.0}
  & 47.2 & \textbf{43.5} & 38.3 & 27.2 & 43.9
  & 41.9 & 40.7 & 29.9 & 35.6 & 35.8
  & \textbf{51.2} & \textbf{49.0} & 44.4
  & 42.1 \\
mContriever
  & 52.5 & 50.1 & 36.4 & 41.8 & 21.5
  & 60.2 & 31.4 & 28.6 & 39.2 & 42.4
  & 48.3 & 39.1 & \textbf{56.0} & 52.8 & 51.7
  & 41.0 & 40.8 & 41.5
  & 43.1 \\
mColBERT
  & {57.1} & 54.6 & 38.8 & 42.6 & 46.0
  & 46.5 & 26.7 & 47.0 & 29.8 & \textbf{49.6}
  & 48.7 & \textbf{47.7} & 35.8 & 46.2 & 48.1
  & 39.8 & 33.4 & \textbf{56.1}
  & 44.1 \\
HyDE-8B
  & 33.8 & 40.5 & 31.1 & 25.6 & 40.2
  & \textbf{65.2} & 20.5 & 33.8 & 49.3 & 29.8
  & 35.0 & 47.6 & 22.9 & \textbf{58.5} & \textbf{58.1}
  & 34.8 & 24.6 & 44.6
  & 38.7 \\
\midrule
STORM\textsubscript{32}	&\uline{55.8}&	\uline{50.1}&	\uline{51.2}	&\uline{40.4}&	\uline{38.2}	&\uline{58.7}	&\uline{31.9}	&\uline{48.3}	&\uline{49.6}&	\uline{42.8}	&\uline{50.0}&	\uline{43.9}	&\uline{44.0}&	\uline{47.5}	&\uline{51.8}	&\uline{32.1}&	\uline{33.9}&	\uline{55.0}&	{45.8}\\
STORM\textsubscript{64}	&\uline{\textbf{57.4}}	&\uline{55.1} &	\uline{51.7}	&\uline{42.5}&	\uline{38.7}&	\uline{61.3}	&\uline{33.4}&	\uline{52.1}&	\uline{50.6}&	\uline{44.8}	&\uline{\textbf{51.4}}	&\uline{46.6}&	\uline{44.9}	&\uline{53.2}	&\uline{54.8}&	\uline{34.3}&	\uline{36.8}	&\uline{52.2}&	{47.9} \\
STORM\textsubscript{128}	&\uline{\textbf{57.4}}&\uline{\textbf{58.6}}	&\uline{\textbf{52.3}}	&\uline{43.2}	&\uline{39.4}&	\uline{61.8}&	\uline{33.3}&	\uline{\textbf{52.7}}	&\uline{\textbf{51.4}}	&\uline{45.7}&	\uline{51.2}&	\uline{47.1}	&\uline{43.4}&	\uline{54.7}&	\uline{56.5}&	\uline{33.6}	&\uline{37.5}&	\uline{50.8}&	{\textbf{48.4}}\\
  %& 56.7 & \textbf{56.1} & \textbf{52.3} & 42.0 & 37.8 & 61.0 & 31.5 & \textbf{52.9} & \textbf{51.6} & 42.5 & \textbf{52.8} & 46.6 & 44.2 & 53.1 & 49.8 & 31.6 & 37.8 & 51.6& \textbf{47.3} \\
\bottomrule
\end{tabular}}
\caption{nDCG@10 on the MIRACL multilingual benchmark across 18 languages.
STORM-8B (with 32, 64 and 128 max tokens), trained exclusively on English MS-MARCO data, surpasses dedicated multilingual dense retrievers on average.
\textbf{Bold}: best overall. \uline{Underline}: statistically significant improvement over BM25 ($p<0.05$).}
\label{tab:miracl}
\end{table*}

\subsection{ Zero-Shot Multilingual (MIRACL)}
\label{sec:miracl}

We evaluate STORM-8B in a zero-shot setting on 18 MIRACL languages, despite training exclusively on English MS-MARCO. Table~\ref{tab:miracl} reports the results. Two findings stand out: STORM generalizes across languages it never saw during finetuning, and it does so while outperforming dense multilingual baselines on average.

Our best model (STORM with up to 128 generated tokens) achieves an
average nDCG@10 of 48.4, a gain of +9.9 points over BM25 and +4.3
points over mColBERT, the strongest dense multilingual baseline.
This is notable because STORM relies on a purely lexical retrieval
backend, while mDPR, mContriever, and mColBERT are all dense retrievers
trained with multilingual supervision. While STORM leads on average,
the dense retrievers---mDPR in particular---remain stronger on several
languages (e.g., Spanish, Persian, Chinese, and German), where semantic
matching outweighs lexical expansion.

STORM improves over BM25 in every language, with the largest gains
where BM25 is weakest---French (+15.0), Chinese (+15.6), and German
(+14.9)---and consistent gains on morphologically complex languages
such as Korean (+9.3) and Finnish (+6.7). Rather than memorizing
English-specific associations, the policy appears to have learned a
language-independent preference for discriminative, content-bearing
terms that Qwen3's multilingual tokenizer realizes in the target
language. Effectiveness grows with the generation budget, from 45.8
(STORM\textsubscript{32}) to 48.4 (STORM\textsubscript{128}).
%% ─────────────────────────────────────────────────────────────────
\subsection{Comparison with GPT-4.1 as Backbone}
\label{sec:gpt}
 
The baselines in Section~\ref{sec:main} all share the same Qwen3 backbone as STORM, ensuring architectural parity. 
To situate STORM within the broader literature, where many published methods rely on much larger proprietary models, Table~\ref{tab:gpt}
compares STORM-8B against eight GPT-4.1-based query rewriters from the reproducibility study of~\citeauthor{Bigdeli2026ARS}, evaluated
on three in-domain sets (DL'19, DL'20, DL-HARD) and five BEIR datasets
(SciFact, TREC-COVID, FiQA, DBpedia-Entity, TREC-News).
 
Despite an order-of-magnitude gap in parameter count, STORM-8B achieves an average nDCG@10 score of 54.8, the highest in Table~\ref{tab:gpt}, ahead of the two strongest GPT-4.1 baselines, MuGI-GPT4 (54.5) and Q2D-GPT4 (54.1).
STORM-8B is particularly strong on TREC-COVID (+4.3 points over GenQRE), which rewards
vocabulary-level lexical coverage rather than semantic paraphrase.
These results suggest that the reward signal used during training
encourages more retrieval-effective expansion than chain-of-thought
prompting of far larger models.
 
% \begin{table}[t]
% \centering
% \tiny
% \setlength{\tabcolsep}{4pt}
% \begin{tabular}{lccc|cccccc|c}
% \toprule
% \multirow{2}{*}{Method} 
% & \multicolumn{3}{c|}{In-domain} 
% & \multicolumn{6}{c|}{Out-of-domain} 
% & \multirow{2}{*}{Avg.} \\
% \cmidrule(lr){2-4} \cmidrule(lr){5-10}
% & dl19 & dl20 & dlhard 
% & SciF & Argu & covid & FiQA & DBP & News 
% & \\
% \midrule
% BM25 & 50.6 & 48.0 & 28.5 & 67.9 & 39.7 & 59.5 & 23.6 & 31.8 & 39.5 & 42.1 \\
% RM3 & 52.2 & 49.0 & 25.1 & 64.6 & 28.6 & 59.3 & 19.2 & 30.8 & 42.6 & 41.2 \\
% GenQr & 54.8 & 53.7 & 29.2 & 72.6 & 40.6 & 68.7 & 23.0 & 34.4 & 46.5 & 47.1 \\
% GenQer & 55.9 & 55.3 & 27.0 & 72.5 & 40.7 & 75.3 & 23.9 & 36.0 & 48.6 & 48.3 \\
% QA-Ex & 68.3 & 64.2 & 30.2 & 70.6 & 39.7 & 70.7 & 26.4 & 37.0 & 45.0 & 50.2 \\
% Q2K & 59.4 & 57.6 & 34.5 & 70.9 & 40.6 & 71.5 & 26.9 & 37.8 & 46.3 & 49.5 \\
% Q2D & 68.7 & 66.2 & 35.0 & 72.0 & 39.7 & 74.3 & 26.0 & 40.6 & 49.8 & 52.5 \\
% LameR & 63.7 & 65.3 & 35.5 & 72.5 & 41.2 & 70.2 & 26.2 & 39.9 & 48.0 & 51.4 \\
% MUGI & 69.5 & 65.8 & 36.5 & 73.5 & 37.6 & 71.4 & 26.4 & 41.0 & 51.6 & 52.6 \\
% CSQE & 69.0 & 65.5 & 36.6 & 72.1 & 39.8 & 69.9 & 24.7 & 39.0 & 47.9 & 51.6 \\
% \midrule
% STORM8 & 68.1 & 65.5 & 33.5 & 73.2 & 44.8 & 77.1 & 28.9 & 39.0 & 48.6 & 53.2 \\
% \bottomrule
% \end{tabular}
% \caption{}
% \label{tab:ndcg10_indomain_outdomain_storm8}
% \end{table}
\begin{table}[H]
  \centering
  {\fontsize{8pt}{9pt}\selectfont

  \setlength{\tabcolsep}{1.pt}
  
\begin{tabular}{lcccccccc|c}
    \toprule
    Method & DL'19 & DL'20 & DL HD & SciF & Covid & FiQA & DBP & News & Avg. \\
    \midrule
    RM3          & 52.2 & 49.0 & 25.1 & 64.6 & 59.3 & 19.2 & 30.8 & 42.6 & 42.9 \\
    GenQR        & 54.8 & 53.7 & 29.2 & 72.6 & 68.7 & 23.0 & 34.4 & 46.5 & 47.9 \\
    GenQRE      & 55.9 & 55.3 & 27.0 & 72.5 & 75.3 & 23.9 & 36.0 & 48.6 & 49.3 \\
    QA-Expa      & 68.3 & 64.2 & 30.2 & 70.6 & 70.7 & 26.4 & 37.0 & 45.0 & 51.6 \\
    Q2K          & 59.4 & 57.6 & 34.5 & 70.9 & 71.5 & 26.9 & 37.8 & 46.3 & 50.6 \\
    Q2D (ZS)     & 68.7 & 66.2 & 35.0 & 72.0 & 74.3 & 26.0 & 40.6 & 49.8 & 54.1 \\
    LameR        & 63.7 & 65.3 & 35.5 & 72.5 & 70.2 & 26.2 & 39.9 & 48.0 & 52.7 \\
    MUGI         & \textbf{69.5} & 65.8 & 36.5 & \textbf{73.5} & 71.4 & 26.4 & \textbf{41.0} & \textbf{51.6} & 54.5 \\
    CSQE         & 69.0 & 65.5 & \textbf{36.6} & 72.1 & 69.9 & 24.7 & 39.0 & 47.9 & 53.1 \\
    \midrule
    \textbf{Ours} & \uline{66.1} & \textbf{\uline{67.3}} & \uline{34.2} & \uline{72.8} & \textbf{\uline{79.6}} & \textbf{\uline{28.4}} & \uline{40.1} & \uline{50.2} & \textbf{\uline{54.8}} \\
    \bottomrule
  \end{tabular}
  }
  \caption{Comparison with GPT-4.1 based rewriters. nDCG@10 on
           in-domain (DL'19, DL'20, DL HD) and five zero-shot BEIR datasets (SciF, COVID, FiQA, DBP, News). Ours: STORM-8B with 64 max tokens uses a Qwen3-8B backbone, while all other rows use GPT-4.1. \textbf{Bold}: best overall; \uline{Underline}: statistically significant improvement over RM3.}
  \label{tab:gpt}
\end{table}

%\subsection{Analysis}
    
\subsection{Efficiency Analysis}
\label{sec:efficiency}
Table~\ref{tab:dl20_robust04_efficiency} reports generation and retrieval latency per query for in-domain and out-of-domain datasets. Among LLM-based rewriters, effectiveness typically comes at the cost of longer prompts or outputs: W2P relies on a 405-token prompt, producing 829 tokens per query and incurring 34.2\,s of generation latency; MuGI generates multiple pseudo-references totalling 521 tokens and 16.8\,s per query. More generated tokens translate into more query terms issued to the index, which increases query processing cost.
STORM departs from this pattern in two ways. First, its short prompt and focused keyword output hold generation latency to 4.39\,s at 8B, dropping to under 1\,s at 0.6B (Table~\ref{tab:qwen3_latency_ndcg_by_size}). This is not incidental: as described in Section~\ref{sec:reward}, the reward explicitly penalizes high document-frequency terms and uses log-likelihood as a length regularizer, directly discouraging the model from generating long or uninformative sequences during training. Second, and more notably, STORM's retrieval latency matches plain BM25 (0.01\,s) despite querying the index with up to 8× more terms (33 vs.\ 4 on DL'20, Table~\ref{tab:dl20_robust04_efficiency}), because the frequency penalty steers the model away from terms with costly index lookups. For example, on the query \textit{``What research is ongoing for new fuel sources''} (Appendix Table~\ref{tab:appendix_fuel_sources_full_case_study}), STORM\textsubscript{32} expands the query from 6 to 21 distinct terms, yet its retrieval still takes only 44.6\,ms---even slightly faster than the original 6-term BM25 query (46.8\,ms)---while raising nDCG@10 from 0.0 to 78.3. Appendix Table~\ref{tab:appendix_bronchioles_full_case_study} reports a similar in-domain example.
Finally, Figure~\ref{fig:latency_ndcg} illustrates this trade-off: MuGI and W2P achieve competitive nDCG@10 but at latencies exceeding 10\,s per query, while HyDE remains fast but sacrifices effectiveness. STORM occupies the upper-left region of the plot---high nDCG@10 at low latency---across all model sizes.

\begin{table}[h]
\centering
\setlength{\tabcolsep}{1.5pt}
{\fontsize{7pt}{5.8pt}\selectfont

\resizebox{\columnwidth}{!}{%
\begin{tabular}{@{}lcc|cc|cc|cc|cc|cc@{}}
\toprule
& \multicolumn{2}{c}{\textbf{BM25}}
& \multicolumn{2}{c}{\textbf{HyDE}}
& \multicolumn{2}{c}{\textbf{MuGI}}
& \multicolumn{2}{c}{\textbf{W2P}}
& \multicolumn{2}{c}{\textbf{STORM\textsubscript{64}}}
& \multicolumn{2}{c}{\textbf{STORM\textsubscript{32}}} \\
\cmidrule(lr){2-3}
\cmidrule(lr){4-5}
\cmidrule(lr){6-7}
\cmidrule(lr){8-9}
\cmidrule(lr){10-11}
\cmidrule(l){12-13}
& DL'20 & RB04
& DL'20 & RB04
& DL'20 & RB04
& DL'20 & RB04
& DL'20 & RB04
& DL'20 & RB04 \\
\midrule
nDCG@10
& 48.0 & 40.8
& 40.7 & 41.4
& 61.7 & 48.4
& 59.9 & 50.4
& \textbf{67.3} & \textbf{56.0}
& 66.2 & 55.9 \\
\midrule
\multicolumn{13}{@{}l}{\textit{Latency (s)}} \\
Generation
& $-$  & $-$
& 4.9  & 7.6
& 16.8 & 21.4
& 34.2 & 39.3
& 4.39 & 4.50
& \textbf{2.31} & \textbf{2.39} \\
Retrieval
& \textbf{0.01} & \textbf{0.01}
& 0.20 & 0.24
& 0.25 & 0.28
& 0.16 & 0.20
& \textbf{0.01} & \textbf{0.01}
& \textbf{0.01} & \textbf{0.01} \\
\midrule
\multicolumn{13}{@{}l}{\textit{Token budget}} \\
Prompt tokens
& $-$  & $-$
& \textbf{21} & \textbf{21}
& 50   & 50
& 405  & 405
& 29   & 29
& 29   & 29 \\
Total tokens/q
& \textbf{7} & \textbf{19}
& 149  & 230
& 521  & 635
& 829  & 891
& 172  & 169
& 96   & 96 \\
BM25 terms/q
& \textbf{4} & \textbf{10}
& 60   & 95
& 101  & 135
& 118  & 141
& 33   & 36
& 20   & 23 \\
\bottomrule
\end{tabular}}
}
\caption{Efficiency and effectiveness comparison on TREC DL'20 and Robust04.
\textit{Generation latency}: average LLM inference time per query in seconds,
reflecting the cost of producing the expanded query.
\textit{Retrieval latency}: average BM25 index query time per query in seconds.
\textit{Prompt tokens}: number of tokens in the input context fed to the LLM.
\textit{Total tokens/q}: average number of tokens in the query issued to retrieval (generated terms for rewriters; the original query for BM25).
\textit{BM25 terms/q}: average number of distinct terms in the BM25 query issued to the
index; larger values increase posting-list traversals and slow retrieval.
Best nDCG@10 in \textbf{bold}.}
\label{tab:dl20_robust04_efficiency}
\end{table}

%% ─────────────────────────────────────────────────────────────────
\subsection{Ablation Study}
\label{sec:ablation}
%% ─────────────────────────────────────────────────────────────────
Tables~\ref{tab:training_ablation_qwen3_4b} and~\ref{tab:inference_ablation_storm_grouped_size} (in Appendix) summarize training and inference ablations respectively.

\paragraph{Model size.}
Comparing rows within each Qwen3 family in Table~\ref{tab:main_results},
STORM scales predictably with parameter count: mean OoD nDCG@10 increases
monotonically from 43.1 (0.6B) to 47.5 (8B), with diminishing returns
at larger sizes ($\Delta{=}{+}1.2$ from 1.7B to 4B,
$\Delta{=}{+}0.9$ from 4B to 8B).

\paragraph{System prompt selection.}
We evaluated 150 candidate prompts on 100 held-out MS-MARCO queries
(Qwen3-0.6B, reward-guided beam search) and selected the highest-reward
one for all experiments; it is reported in Appendix~\ref{sec:prompt}.

\paragraph{Generation temperature.}
We ablate beam-search temperature $\tau \in \{0, 0.5, 1.0\}$.
Greedy beam search ($\tau{=}0$) yields little improvement over the
base model, as the lack of stochasticity limits exploration and leads
to repetitive expansions.
Both $\tau{=}0.5$ and $\tau{=}1.0$ improve retrieval quality and
reduce overfitting over longer training runs. $\tau{=}1.0$ is selected for its higher beam diversity.
 
\paragraph{Mixture weight $\varepsilon$.}
We sweep $\varepsilon \in \{0.1, 0.2, 0.3\}$.
Setting $\varepsilon{=}0.3$ slows convergence, as too many samples
come from unconstrained sampling rather than the reward-guided beam.
$\varepsilon{=}0.1$ and $\varepsilon{=}0.2$ achieve comparable final
performance; we select $\varepsilon{=}0.2$ as it provides a slightly
more robust training signal.

\paragraph{Beam selection strategy.}
Section~\ref{sec:method} selects the highest-reward beam from the
$K$ beams; we ablate this against uniform random selection over
the beam set. Both yield similar asymptotic performance, but
best-beam selection accelerates early learning.
 
\paragraph{Number of beams $K$.}
We fix $K \in \{2, 5, 10\}$.
$K{=}2$ plateaus early due to limited diversity;
$K{=}10$ produces noisier reward estimates.
$K{=}5$ provides the best balance and is used throughout.
 
\paragraph{LoRA vs.\ full fine-tuning.}
As shown in Table~\ref{tab:inference_ablation_storm_grouped_size},
LoRA already yields strong performance, and increasing the rank
consistently improves results. Full fine-tuning provides a further
gain, though modest: the gap is largest at 0.6B (${\approx}2$
nDCG@10 points) and negligible at 8B (0.4 points), suggesting
that full fine-tuning matters most when model capacity is limited.

\paragraph{Maximum generation length.}
We compare generation budgets of 32, 64, and 128 tokens
(Table~\ref{tab:inference_ablation_storm_grouped_size}).
For smaller models (0.6B and 1.7B), 32 tokens are sufficient and
yield similar or better results than longer budgets, likely because
smaller models do not benefit from generating additional terms.
For larger models, longer budgets help, particularly on multilingual
benchmarks (MIRACL) where 128 tokens yield the best results, while
64 tokens offer the best latency--effectiveness trade-off for
in-domain and out-of-domain queries.

\paragraph{Additional design choices.}
Replacing Hard nDCG with SoftNDCG~\cite{satouf-etal-2026-quester}
slowed learning without improving final performance.
Substituting sparse MS-MARCO labels with cross-encoder pseudo-labels
consistently improved the training signal.
Increasing retrieval depth from top-100 to top-500 was not explored
due to the substantial increase in runtime; we therefore retain top-100.

\section{Conclusion}
\label{sec:conclusion}

We presented \textbf{STORM}, a query rewriting framework that turns the retrieval reward into a token-level signal through reward-guided beam search and trains the policy on the selected rewrites in a self-supervised with retrieval feedback manner, with no human-written rewrites. STORM substantially improves retrieval effectiveness over both prompting-based and reinforcement-learning rewriters, yet remains \emph{infrastructure-light}: it retrieves as fast as plain BM25 on a standard inverted index that need not be rebuilt as models evolve. These gains are robust across scale---0.6B backbones already surpass their same-size baselines, while at 8B STORM rivals rewriters built on far larger proprietary models---and transfer zero-shot to 18 languages, where STORM outperforms dedicated multilingual dense retrievers on average. Together, these results position lexical query rewriting as a competitive and efficient alternative to dense neural retrieval, and suggest reward-guided decoding as a general tool for aligning generation with non-differentiable retrieval objectives.

\section{Limitations}
\label{sec:limitations}
%\paragraph{Training-time scalability.} At every generation step, STORM issues a BM25 query for each of the $K$ partial candidates, so training cost scales with both the beam width and the  corpus size. With a low beam size (e.g. 5) on MS-MARCO v1 (8.8M passages) this remains tractable, but the overhead would grow on web-scale corpora or with larger beams. Future work should focus on an incremental scoring procedure that reuses the parent beam's retrieved documents, reducing the marginal cost of each expansion step.
\paragraph{In-domain effectiveness gap.}
On in-domain benchmarks, STORM closes a large fraction of the gap to SPLADE-v2 compared with other query rewriting baselines, but still does not match it; on out-of-domain BEIR collections, our method can achieve higher performance. This pattern suggests that STORM effectively learns discriminative lexical terms for retrieval. However, it remains limited in capturing the richer contextual and semantic relationships encoded by learned sparse retrievers through their term-weighting mechanisms.
\paragraph{Generation length sensitivity.}
The optimal maximum generation length is dataset dependent. For example, 64 tokens are sufficient for many in-domain and out-of-domain queries, whereas 128 tokens give the best results on MIRACL. Choosing a single fixed budget therefore involves a trade-off between latency and effectiveness, and we leave adaptive or query-dependent length control to future work.
%\end{document}

% Bibliography entries for the entire Anthology, followed by custom entries
%\bibliography{anthology,custom}
% Custom bibliography entries only
\bibliography{custom}
\newpage
\appendix

\section{ Appendix}
\label{sec:appendix}

\subsection{Training Details}
\label{sec:training_details}

\paragraph{Training data.} We sample ${\approx}80$k queries from the MS-MARCO passage-retrieval training set~\cite{craswell2021msmarco,bajaj2016msmarco} to train our models. We use BM25 as the retriever via Pyserini~\cite{lin2021pyserinieasytousepythontoolkit}, which is used to compute the reward function described in Section~\ref{sec:reward}. To monitor training progress, we hold out a set of 500 queries that are excluded from optimization and used for checkpoint selection.

\paragraph{Optimization.} We instantiate STORM with four sizes of the Qwen3 family~\cite{qwen3technicalreport}: 0.6B, 1.7B, 4B, and 8B parameters. To reduce computational cost, we disable Chain-of-Thought reasoning and constrain the maximum generation length to 32 tokens. The models are trained for one epoch using AdamW, with a learning rate of $10^{-6}$, a batch size of 32, and gradient accumulation over 8 steps. During training, the reward-guided beam component uses beam width $K{=}5$, while the stochastic component (nucleus sampling) uses standard sampling. Both generation methods share the same temperature, top-$k$, and top-$p$ settings, ensuring that the mixture proposal in Eq.~\ref{eq:proposal} is defined consistently. Experiments were conducted on NVIDIA RTX A6000, H100, and H200 GPUs. Depending on model size, each run required between 2 and 5 days of training.\footnote{Training is implemented using \href{https://huggingface.co}{HuggingFace}.}\footnote{\href{https://openai.com}{OpenAI} \& \href{https://claude.ai}{Claude AI} were used to accelerate code development.}

\paragraph{Beam generation tokens during training}
The maximum generation length is an upper bound rather than a required output
length: generation may terminate earlier through EOS. During training,
reward-guided rewrites have a median length of 11 tokens and a mean of 12.5
tokens, with 90\% shorter than 25 tokens, despite the 32-token cap. Thus the
cap bounds exploration but does not force the policy to exhaust the available
budget.

\begin{figure*}[t]
    \centering
    \includegraphics[width=\linewidth]{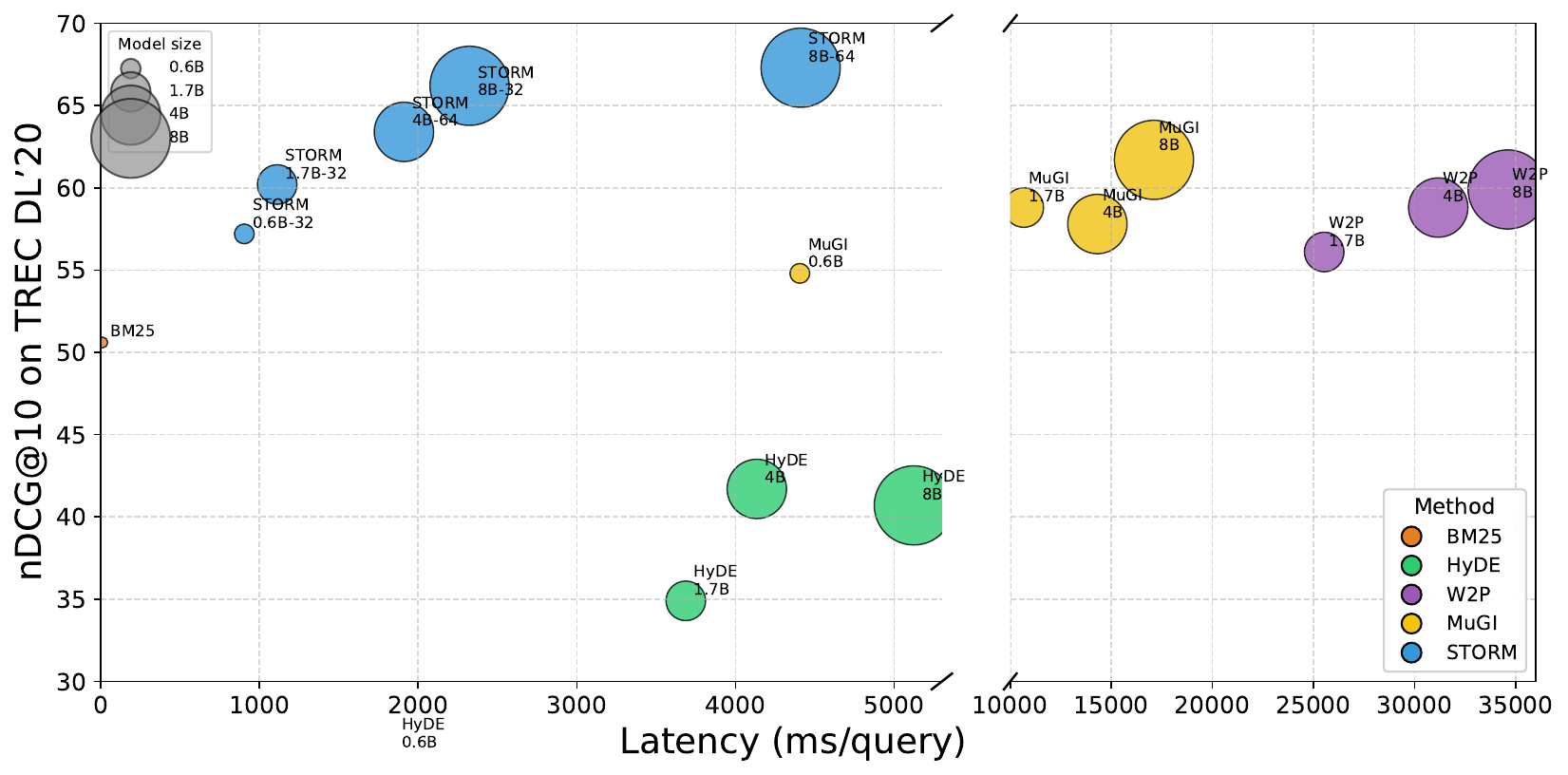}
   \caption{
Latency--effectiveness comparison on TREC DL'20.
The x-axis shows total per-query latency (generation + retrieval) in
milliseconds, with a broken scale to accommodate the wide latency
range across methods. The y-axis shows nDCG@10. Bubble sizes encode
model size, colors indicate method families, and labels report the
backbone size and (for STORM) the maximum generation length.
STORM occupies the upper-left region of the plot, achieving the
highest nDCG@10 at substantially lower latency than competing
LLM-based rewriters. Two points fall outside the plotted range:
W2P-0.6B (47\,542\,ms/query), whose latency is inflated by repeated
generation attempts after parsing failures, and HyDE-0.6B
(nDCG@10$=$25.6), which falls below the y-axis floor.
}
    \label{fig:latency_ndcg}
\end{figure*}

\begin{table*}[h]
\centering
\small
\setlength{\tabcolsep}{3pt}
\begin{tabular}{c l c c c c c c c c c c}
\hline
\# & Training & $\varepsilon$ & LR & Temp. & Sample & Top-$k$/beam & $\lambda_{L}$ & $\lambda_{\mathrm{DF}}$ & Best beam & Score-t & Train loss \\
\hline
1  & Full & 0.2 & $1{\times}10^{-6}$ & 1.0 & true  & 5 & 0.01  & 0.005  & true  & \textbf{0.4820} & 0.2408 \\
2  & LoRA $r{=}64$ & 0.1 & $1{\times}10^{-5}$ & 1.0 & true  & 5 & 0.01  & 0.005  & false & 0.4702 & 0.3336 \\
3  & LoRA $r{=}64$ + SPLADE signal & 0.1 & $1{\times}10^{-5}$ & 1.0 & true  & 5 & 0.01  & 0.005  & false & {0.4790} & 0.2874 \\
4  & Full + SPLADE signal & 0.1 & $1{\times}10^{-5}$ & 1.0 & true  & 5 & 0.01  & 0.005  & false & 0.4735 & 0.3239 \\
5  & LoRA $r{=}64$ + SPLADE signal & 0.1 & $1{\times}10^{-5}$ & 1.0 & true  & 5 & 0.01  & 0.005  & true  & 0.4628 & 0.1978 \\
6  & LoRA $r{=}64$ & 0.1 & $1{\times}10^{-5}$ & 1.0 & true  & 5 & 0.01  & 0.005  & false & 0.4703 & 0.3902 \\
7  & LoRA $r{=}64$ & 0.1 & $1{\times}10^{-5}$ & 1.0 & true  & 5 & 0.001 & 0.0005 & false & 0.4578 & 1.1033 \\
8  & LoRA $r{=}64$ & 0.1 & $1{\times}10^{-5}$ & 1.0 & true  & 2  & 0.01  & 0.005  & false & 0.4727 & 0.3063 \\
9  & LoRA $r{=}64$ & 0.1 & $1{\times}10^{-5}$ & 0.5 & true  & 5  & 0.01  & 0.005  & false & {0.4741} & 0.1084 \\
10 & LoRA $r{=}64$ & 0.1 & $1{\times}10^{-5}$ & 1.0 & true  & 10 & 0.01  & 0.005  & false & 0.4687 & 0.3244 \\
11 & LoRA $r{=}64$ & 0.1 & $1{\times}10^{-5}$ & 1.0 & true  & 5  & 0.01  & 0.005  & true  & 0.4668 & 0.2425 \\
12 & LoRA $r{=}64$ & 0.1 & $1{\times}10^{-5}$ & 1.0 & true  & 5  & 0.1   & 0.05   & false & 0.4726 & \textbf{0.0284} \\
13 & LoRA $r{=}64$ & 0.3 & $1{\times}10^{-5}$ & 1.0 & true  & 5  & 0.01  & 0.005  & false & 0.4664 & 0.3570 \\
14 & LoRA $r{=}64$ & 0.1 & $1{\times}10^{-5}$ & 0.0 & false & 5  & 0.01  & 0.005  & false & 0.4579 & 0.0515 \\
\hline
\end{tabular}
\caption{
Ablation study of the training configuration for the Qwen3-4B model.
All runs are trained for 50 optimization steps with an effective batch size of 256, corresponding to batch size 32 and 8 gradient-accumulation steps.
Columns follow the notation of Section~\ref{sec:method}: $\varepsilon$ is the mixture weight of Eq.~\ref{eq:proposal}, LR the learning rate, Temp.\ the beam-search temperature $\tau$, \emph{Sample} whether stochastic (nucleus) sampling is enabled, Top-$k$/beam the beam width $K$, and $\lambda_{L}$/$\lambda_{\mathrm{DF}}$ the log-likelihood and document-frequency penalty weights of the reward (Section~\ref{sec:reward}). \emph{Best beam} indicates whether the highest-reward beam is selected (vs.\ uniform random selection).
We report the held-out MS-MARCO 500 mean target score, denoted Score-t, and the training loss.
The direct SPLADE retrieval/passage-generation run is excluded to keep the comparison focused on the BM25 keyword-generation setting.
}
\label{tab:training_ablation_qwen3_4b}
\end{table*}

\begin{table*}[t]
\centering
\small
\setlength{\tabcolsep}{3.5pt}
\resizebox{\textwidth}{!}{
\begin{tabular}{llccc|cccccccccc|c}
\toprule
\textbf{Size} & \textbf{Method}
& \textbf{DL-19} & \textbf{DL-20} & \textbf{Dev}
& \textbf{NFC} & \textbf{SCID} & \textbf{SciF} & \textbf{Touché} & \textbf{FiQA} & \textbf{Covid}
& \textbf{Signal} & \textbf{News} & \textbf{RB04} & \textbf{DBP}
& \textbf{Avg.} \\
\midrule

\multirow{3}{*}{8B}

& STORM\textsubscript{128}&
67.4	&\textbf{68.2}	&\textbf{22.9}&	36.2	& 15.7&	72.1&	45.4	&\textbf{28.4}&	\textbf{79.9}&	33.9&	\textbf{50.3}&	55.1&	39.5&45.7\\

& STORM\textsubscript{64}
& 66.1 & {67.3} & \textbf{22.9}
& 36.8 & \textbf{15.8} & \textbf{72.8} & 44.4 & \textbf{28.4} & {79.6}
& 33.5 & {50.2} & \textbf{56.0} & \textbf{40.1}
& \textbf{45.8} \\

& STORM\textsubscript{32}
& 65.7 & 66.2 & 22.6
& \textbf{37.0} & \textbf{15.8} & 72.2 & 44.6 & 28.1 & 77.7
& 33.0 & 48.9 & 55.9 & 39.3
& 45.2 \\

& STORM-LoRA\textsubscript{64}
& {67.9} & 66.4 & 22.8
& 36.7 & \textbf{15.8} & 71.9 & 46.1 & 28.6 & 77.1
& \textbf{34.4} & 49.6 & 53.9 & 39.7
& 45.4 \\
& STORM\textsubscript{32} Greedy &
65.6&	63.8	&21.6	&36.8&	15.5&	71.2&	43.8&	27.3&	73.1&	32.5&	47.1	&55.8&	38.3&	44.1\\
& STORM\textsubscript{64} Greedy &
\textbf{68.0}&	65.9	&22.2&	36.3&	15.5&	71.8&	45.0	&27.4&	76.3	&32.6&	48.5	&54.9	&38.8&	44.7\\
\midrule

\multirow{2}{*}{4B}
& STORM\textsubscript{64}
& 63.7 & 64.4 & 22.3
& 36.4 & 15.5 & 72.2 & 47.3 & 27.7 & 75.2
& 33.6 & 48.7 & 53.5 & 39.6
& 45.0 \\

& STORM\textsubscript{32}& 62.4	&63.3&	22.7&	36.1&	15.6	&70.8	&46.2&	27.7	&73.4	&32.6	&48.1&	52.9&	39.1&	44.3 \\

& STORM-LoRA\textsubscript{64}
& 62.7 & 64.3 & 22.7
& 36.5 & 15.6 & 71.6 & \textbf{47.6} & 27.7 & 74.5
& 31.2 & 46.3 & 53.0 & 38.7
& 44.3 \\

\midrule

\multirow{2}{*}{1.7B}
& STORM\textsubscript{32}
& 63.3 & 60.2 & 21.7
& 36.2 & 15.3 & 71.6 & 47.3 & 27.9 & 75.2
& 32.8 & 45.7 & 52.2 & 37.4
& 44.2 \\

& STORM\textsubscript{64}&
61.9	&62.2&	21.6&	35.5	&14.9&	71.3&	47.0	&27.3&	73.6&	32.7&	45.3	&52.4	&37.3&	43.7\\

& STORM-LoRA\textsubscript{64}
& 53.9 & 58.9 & 19.9
& 35.0 & 14.5 & 69.4 & 41.6 & 25.8 & 75.7
& 30.6 & 43.7 & 52.8 & 32.7
& 42.2 \\

\midrule

\multirow{2}{*}{0.6B}
& STORM\textsubscript{32}
& 56.7 & 57.2 & 21.0
& 32.5 & 14.7 & 70.2 & 45.9 & 25.7 & 72.5
& 32.3 & 40.4 & 48.5 & 34.2
& 41.7 \\

&STORM\textsubscript{64}& 54.2&	54.2	&20.4&	32.0&14.2&	70.1	&44.1	&24.9&	68.6&	30.9&	39.0	&46.9	&32.9	&40.4\\

& STORM-LoRA\textsubscript{64}
& 52.1 & 50.2 & 19.0
& 32.8 & 14.1 & 67.0 & 42.8 & 23.5 & 66.7
& 31.9 & 41.6 & 44.0 & 33.0
& 39.7 \\

\bottomrule
\end{tabular}
}
\caption{
Inference-time ablation of STORM query specification generation,
grouped by Qwen3 backbone size.
We compare STORM variants across backbone size, LoRA/full fine-tuning
setting, maximum generation length (32, 64, or 128 tokens), and
decoding strategy: diverse beam search with 3 returned sequences (default)
versus greedy decoding with a single sequence.
DL-19 and DL-20 report nDCG@10, Dev reports the corresponding MS-MARCO
development score, and out-of-domain effectiveness is reported across
ten BEIR datasets (HotpotQA and NQ are omitted here, so the average is
not directly comparable to the 12-dataset Avg.\ OoD of Table~\ref{tab:main_results}).
The Avg. column reports the macro-average over these ten out-of-domain datasets.
Best values are shown in bold.
}
\label{tab:inference_ablation_storm_grouped_size}
\end{table*}

\begin{table*}[t]
\centering
\small
\setlength{\tabcolsep}{6pt}
\begin{tabular}{llccc}
\toprule
\textbf{LLM size} & \textbf{Method}
& \textbf{Generation} & \textbf{Retrieval} & \textbf{nDCG@10} \\
 & & \textbf{ms/query} & \textbf{ms/query} & \\
\midrule
-- & BM25 & -- & 10.7 & 50.6 \\
\midrule
\multirow{5}{*}{8B}
& HyDE  & 4912.4  & 210.7 & 40.7 \\
& MuGI  & 16856.6 & 251.2 & 61.7 \\
& W2P   & 34437.4 & 160.8 & 59.9 \\
& STORM$_{\text{64}}$ & 4396.0 & 14.7 & \textbf{67.3} \\
& STORM$_{\text{32}}$ & 2309.8 & 14.3 & 66.2 \\
\midrule
\multirow{4}{*}{4B}
& HyDE  & 4018.5  & 115.4 & 41.7 \\
& MuGI  & 14162.4 & 146.9 & 57.8 \\
& W2P   & 31073.4 & 91.9  & 58.8 \\
& STORM$_{\text{32}}$ & 1896.4 & 13.1 & 63.4 \\
\midrule
\multirow{4}{*}{1.7B}
& HyDE  & 3565.2  & 121.1 & 34.9 \\
& MuGI  & 10517.0 & 152.6 & 58.8 \\
& W2P   & 25405.8 & 119.7 & 56.1 \\
& STORM$_{\text{32}}$ & 1097.8 & 12.4 & 60.2 \\
\midrule
\multirow{4}{*}{0.6B}
& HyDE  & 1553.7 & 60.6 & 25.6 \\
& MuGI  & 3794.2 & 64.0 & 54.8 \\
& W2P   & 47542.3& 67.7 & 51.1 \\
& STORM$_{\text{32}}$ & 890.7 & 13.1 & 57.2 \\
\bottomrule 
\end{tabular}
\caption{
Efficiency--effectiveness comparison of BM25 and LLM-based query expansion methods grouped by Qwen3 model size, in-domain for TREC-DL-2020.
Latency is decomposed into generation and retrieval time, both reported in milliseconds per query. 
All neural methods use Qwen3 models ranging from 0.6B to 8B parameters. 
For STORM, retrieval time is almost as fast as the initial BM25 query. 
STORM$_{\text{64}}$ and STORM$_{\text{32}}$ denote maximum generation budgets of 64 and 32 tokens, respectively. 
The best nDCG@10 score is shown in bold, and all STORM results are
statistically significant improvements over BM25.
}
\label{tab:qwen3_latency_ndcg_by_size}
\end{table*}

\subsection{Prompt format.}
\label{sec:prompt}
We use a single prompt template across training and both in-domain and out-of-domain
evaluations. The model receives the original query and is asked to produce a
compact keyword specification that can be issued directly to the lexical
retriever.

\begin{promptbox}{Standard keyword-generation prompt}
\small

\texttt{From the query generate new semantic related keywords.}\\
\texttt{Output the result strictly as a single comma-separated line.}

\vspace{1mm}
\texttt{[QUERY]: \{query\}}\\
\texttt{[KEYWORDS]:}
\end{promptbox}

For MIRACL, we use the same input--output format, but add a language constraint
to ensure that the generated keywords match the language.

\begin{promptbox}{Multilingual MIRACL prompt}
\small

\texttt{From the query generate new semantic related keywords.}\\
\texttt{Output the result strictly as a single comma-separated line.}\\
\texttt{The keywords must be in \{language\} language.}

\vspace{1mm}
\texttt{[QUERY]: \{query\}}\\
\texttt{[KEYWORDS]:}
\end{promptbox}

\begin{table*}[t]
\centering
\scriptsize
\setlength{\tabcolsep}{3pt}
\renewcommand{\arraystretch}{1.18}
\begin{tabularx}{\textwidth}{p{1.cm} p{7.5cm} r r r r r}
\toprule

\textbf{Query:} 
& \textbf{"what type of tissue are bronchioles"} 
& 
& 
& 
& \\
\toprule
\textbf{Method} 
& \textbf{BM25-analyzed term counts} 
&\textbf{\#Analyzed} 
& \textbf{\#Gen.} 
& \textbf{nDCG@10} 
& \textbf{Recall@1k} 
& \textbf{Retrieve} 

\\
& &\textbf{Unique Terms} &\textbf{ Terms} & & &\textbf{Time(ms)}  \\

\midrule

BM25 
& \texttt{\{what: 1, type: 1, tissu: 1, bronchiol: 1\}}
& 4 
& -
& 44.1 
& 63.6 
& \textbf{25.7}
  \\

\midrule

HyDE 
& \texttt{\{bronchiol: 3, small: 1, airwai: 3, within: 1, respiratori: 1, system: 1, compos: 1, pseudostratifi: 1, ciliat: 1, columnar: 1, epithelium: 1, type: 1, tissu: 2, consist: 1, multipl: 1, layer: 2, cell: 3, appear: 1, stratifi: 1, under: 1, microscop: 1, actual: 1, singl: 1, vari: 1, height: 1, epitheli: 2, cover: 1, cilia: 1, which: 2, help: 1, move: 1, mucu: 1, trap: 1, particl: 1, out: 1, aid: 1, clearanc: 1, debri: 1, pathogen: 1, also: 1, contain: 1, smooth: 1, muscl: 1, allow: 1, them: 1, constrict: 1, dilat: 1, regul: 2, airflow: 1, combin: 1, muscular: 1, enabl: 1, perform: 1, critic: 1, role: 1, ga: 1, exchang: 1\}}
& 57 
& 105
& 13.2 
& 57.6 
& 108.1 \\

\midrule

MuGI 
& \texttt{\{what: 18, type: 23, tissu: 28, bronchiol: 26, small: 5, air: 5, passag: 6, within: 3, lung: 5, compos: 5, pulmonari: 1, specif: 1, pseudostratifi: 5, ciliat: 6, columnar: 6, epithelium: 6, line: 1, cilia: 4, goblet: 2, cell: 10, which: 3, help: 5, trap: 6, move: 5, mucu: 6, particul: 1, matter: 1, out: 3, airwai: 5, prevent: 1, infect: 1, maintain: 3, clear: 2, breath: 1, also: 2, contain: 4, smooth: 4, muscl: 4, allow: 4, control: 1, airflow: 3, through: 1, bronchoconstrict: 4, bronchodil: 3, primarili: 1, epitheli: 2, includ: 1, debri: 2, respiratori: 5, system: 4, wall: 1, regul: 2, overal: 1, structur: 4, design: 1, facilit: 2, ga: 2, exchang: 2, found: 1, known: 1, consist: 3, multipl: 1, layer: 4, appear: 3, stratifi: 3, actual: 1, singl: 2, some: 3, taller: 1, than: 1, other: 1, particl: 3, while: 2, secret: 1, special: 2, crucial: 1, clean: 1, function: 1, due: 2, vari: 2, height: 2, upward: 1, awai: 2, from: 2, underli: 1, dilat: 1, essenti: 1, effici: 2, defens: 2, support: 1, basement: 1, membran: 1, surround: 1, against: 1, pathogen: 1\}}
& 95 
&490
& 48.0 
& 75.0 
& 181.4 \\

\midrule

W2P 
& \texttt{\{type: 142, tissu: 231, bronchiol: 207, epitheli: 15, smooth: 59, connect: 40, ciliat: 21, pseudostratifi: 18, columnar: 18, muscl: 59, epithelium: 21, respiratori: 37, function: 12, ga: 28, compos: 25, line: 12, airflow: 46, exchang: 28, typic: 3, small: 18, thin: 3, wall: 15, branch: 3, bronchial: 3, tree: 3, lead: 3, alveoli: 9, lung: 30, primarili: 6, usual: 3, made: 3, help: 15, clear: 3, mucu: 6, trap: 6, particl: 6, surround: 6, layer: 6, control: 6, diamet: 6, regul: 34, also: 9, contain: 15, elast: 9, fiber: 9, nerv: 3, contribut: 3, flexibl: 3, respons: 6, stimuli: 6, work: 6, togeth: 9, maintain: 15, proper: 9, facilit: 3, contract: 12, airwai: 27, relax: 12, enabl: 3, allow: 15, plai: 12, kei: 3, role: 12, bronchoconstrict: 3, bronchodil: 3, provid: 3, structur: 12, support: 3, ensur: 6, effici: 9, system: 21, constrict: 9, dilat: 9, part: 6, conduct: 3, zone: 3, tract: 3, crucial: 9, cell: 12, variou: 3, featur: 3, make: 3, essenti: 6, goblet: 6, includ: 3, tubular: 3, found: 3, specif: 3, within: 3, known: 3, character: 3, appear: 3, presenc: 3, cilia: 3, movement: 3, aid: 3, bodi: 3, defens: 3, pathogen: 3, air: 3, passag: 3, health: 3\}}
& 102 
&1802
& 53.4 
& 71.2 
& 131.3 \\

\midrule

STORM$_{32}$ 
& \texttt{\{bronchiol: 13, tissu: 10, type: 6, epitheli: 3, connect: 2, respiratori: 2, line: 2, simpl: 3, ciliat: 2, columnar: 3, epithelium: 3, bron: 1, includ: 1, smooth: 1\}}
& 14 
&\textbf{ 54}
& 92.7 
& 76.5 
& 27.1 \\

\midrule

STORM$_{64}$ 
& \texttt{\{bronchiol: 23, tissu: 22, type: 12, epitheli: 3, connect: 3, respiratori: 5, line: 3, simpl: 3, ciliat: 3, columnar: 3, epithelium: 4, includ: 5, smooth: 3, muscl: 3, part: 4, system: 3, wall: 3, basement: 1, membran: 1, endothelium: 1, airwai: 1, consist: 1\}}
& 22 
&123
& \textbf{97.9} 
& \textbf{77.3} 
& 40.1 \\

\bottomrule
\end{tabularx}

\caption{
Qualitative appendix example for the query qid:"914916" from  DL-20 dataset in-domain.
All query representations are shown after applying the same BM25 analyzer used at retrieval time.
Thus, terms appear in their analyzed form, e.g., \texttt{tissu}, \texttt{bronchiol}, \texttt{epitheli}, and \texttt{airwai}.
The table reports the analyzed term-count dictionary, the number of unique analyzed terms, the total number of generated analyzed terms, and the resulting retrieval effectiveness and efficiency.
Compared with HyDE, MuGI, and W2P, STORM produces a much more compact query specification while achieving substantially higher nDCG@10 and Recall@1000, with retrieval time close to the original BM25 query.
}
\label{tab:appendix_bronchioles_full_case_study}
\end{table*}

\begin{table*}[t]
\centering
\scriptsize
\setlength{\tabcolsep}{3pt}
\renewcommand{\arraystretch}{1.18}
\begin{tabularx}{\textwidth}{p{1cm} p{8.5cm} r r r r r}
\toprule
\multicolumn{4}{l}{\textbf{Query:} \emph{``What research is ongoing for new fuel sources.''}} \\
\midrule
\textbf{Method} 
& \textbf{BM25-analyzed term counts} 
& \textbf{\#Analyzed} 
& \textbf{\#Gen.} 
& \textbf{nDCG@10} 
& \textbf{recall@1k} 
& \textbf{Retrieve} \\
& & \textbf{unique terms}& \textbf{terms} & & & \textbf{time(ms)} \\
\midrule

BM25
& \texttt{\{what: 1, research: 1, ongo: 1, new: 1, fuel: 1, sourc: 1\}}
& 6
& -
& 0.0
& 48.2
& 46.8 \\

\midrule

HyDE
& \texttt{\{ongo: 1, research: 2, new: 2, fuel: 4, sourc: 3, focus: 1, find: 1, sustain: 2, effici: 2, environment: 1, friendli: 1, altern: 2, tradit: 2, fossil: 2, on: 1, major: 1, area: 1, develop: 2, biofuel: 2, which: 1, deriv: 1, from: 1, organ: 1, materi: 1, plant: 1, alga: 1, wast: 2, biomass: 1, scientist: 1, explor: 1, wai: 1, optim: 1, product: 2, ethanol: 1, biodiesel: 1, advanc: 3, reduc: 2, cost: 1, increas: 1, energi: 7, output: 1, anoth: 1, signific: 1, direct: 1, hydrogen: 3, work: 1, improv: 1, method: 1, electrolysi: 1, power: 1, renew: 2, better: 1, storag: 3, transport: 2, solut: 1, make: 2, viabl: 1, industri: 1, applic: 1, addition: 1, substanti: 1, interest: 1, nuclear: 2, fusion: 3, potenti: 2, clean: 1, unlik: 1, fission: 1, ha: 1, produc: 1, vast: 1, amount: 1, minim: 1, risk: 1, meltdown: 1, experiment: 2, reactor: 2, like: 2, intern: 1, thermonuclear: 1, iter: 1, progress: 1, achiev: 1, reaction: 1, solar: 1, wind: 1, also: 1, be: 1, combin: 1, technolog: 1, solid: 1, state: 1, batteri: 1, other: 1, system: 1, enhanc: 1, reliabl: 1, innov: 1, crucial: 1, relianc: 1, mitig: 1, impact: 1, climat: 1, chang: 1\}}
& 104
& 202
& 77.3
& 61.2
& 100.0 \\

\midrule

MuGI
& \texttt{\{what: 18, research: 25, ongo: 23, new: 23, fuel: 34, sourc: 25, focus: 5, develop: 5, sustain: 6, effici: 5, altern: 5, fossil: 6, scientist: 5, explor: 5, advanc: 6, biofuel: 5, deriv: 3, from: 3, alga: 5, agricultur: 3, wast: 3, which: 5, offer: 5, higher: 2, energi: 17, yield: 2, lower: 5, emiss: 2, addition: 5, hydrogen: 7, cell: 5, technolog: 2, be: 6, refin: 4, us: 4, transport: 3, industri: 4, effort: 3, improv: 2, storag: 3, product: 3, method: 3, meanwhil: 2, nuclear: 6, fusion: 5, next: 1, gener: 1, reactor: 1, investig: 3, potenti: 4, long: 3, term: 3, solut: 3, innov: 4, aim: 8, reduc: 5, environment: 4, impact: 4, enhanc: 2, secur: 3, worldwid: 1, carbon: 4, footprint: 4, stationari: 1, power: 3, system: 1, emphasi: 2, green: 1, via: 1, renew: 4, also: 3, replic: 2, sun: 2, produc: 2, process: 2, control: 1, environ: 1, while: 2, meet: 1, global: 1, demand: 1, base: 2, biodiesel: 2, cellulos: 2, ethanol: 2, support: 1, transit: 1, more: 1, futur: 1, vehicl: 2, applic: 1, through: 1, like: 2, electrolysi: 1, solar: 2, wind: 2, major: 1, area: 1, studi: 1, provid: 1, nearli: 2, limitless: 2, minim: 1, address: 1, climat: 2, chang: 2, challeng: 1, relianc: 1, tradit: 1, breakthrough: 1, pursu: 1, zero: 1, though: 1, still: 1, experiment: 1, stage: 1, promis: 1, batteri: 1, critic: 1, integr: 1, grid: 1, depend: 1, non: 1, resourc: 1, mitig: 1\}}
& 125
& 621
& 55.5
& 67.1
& 93.8 \\

\midrule

W2P
& \texttt{\{what: 80, research: 138, ongo: 107, new: 117, fuel: 199, sourc: 135, hydrogen: 49, nuclear: 34, sustain: 52, renew: 12, carbon: 21, emiss: 12, altern: 40, cell: 15, biofuel: 52, advanc: 24, fossil: 34, energi: 62, technolog: 30, focus: 15, develop: 21, effici: 24, replac: 3, world: 3, seek: 6, scientist: 15, explor: 15, option: 9, effort: 12, aim: 12, reduc: 21, depend: 6, finit: 6, resourc: 6, addition: 12, like: 6, solar: 6, wind: 6, power: 6, integr: 3, storag: 9, solut: 12, improv: 3, govern: 9, privat: 9, sector: 9, invest: 6, heavili: 6, area: 3, drive: 3, innov: 12, ensur: 3, cleaner: 9, futur: 6, synthet: 25, climat: 18, batteri: 6, environment: 21, chang: 15, impact: 12, enhanc: 12, secur: 21, current: 9, major: 3, focu: 3, scientif: 3, industri: 3, worldwid: 3, varieti: 6, relianc: 12, mitig: 9, us: 3, transport: 3, deriv: 6, alga: 6, agricultur: 3, wast: 3, investig: 12, fusion: 24, captur: 12, pursu: 6, provid: 3, initi: 3, address: 6, long: 6, term: 6, combat: 3, studi: 3, scalabl: 3, viabil: 3, fund: 3, extens: 3, project: 3, acceler: 3, adopt: 3, greenhous: 6, ga: 6, friendli: 6, activ: 3, includ: 6, find: 3, tradit: 6, signific: 3, interest: 3, potenti: 3, decreas: 3, also: 6, conduct: 3, base: 6, direct: 3, air: 3, system: 3, convert: 3, atmospher: 3, dioxid: 3, usabl: 3, environ: 3, field: 3, engin: 3, geotherm: 3, variou: 3, transit: 3, toward: 3\}}
& 123
& 2121
& 50.2
& 68.8
& 83.5 \\

\midrule

\textbf{STORM$_{32}$}
& \texttt{\{fuel: 18, research: 15, new: 11, sourc: 9, energi: 3, ongo: 7, altern: 3, hydrogen: 3, solar: 3, be: 1, develop: 1, biofuel: 2, nuclear: 3, fusion: 3, institut: 1, work: 1, cell: 2, studi: 2, materi: 1, biomass: 1, transport: 1\}}
& 21
& 92
& \textbf{78.3}
& \textbf{69.4}
& \textbf{44.6} \\

\midrule

\textbf{STORM$_{64}$}
& \texttt{\{fuel: 22, research: 22, new: 14, sourc: 13, energi: 5, ongo: 7, altern: 3, hydrogen: 3, solar: 3, be: 1, develop: 1, biofuel: 3, nuclear: 3, fusion: 3, institut: 2, work: 1, cell: 2, studi: 1, materi: 1, project: 1, wind: 1, improv: 1, marin: 1, electr: 1, vehicl: 1, batteri: 1, academ: 1, govern: 1, fund: 1, technolog: 1, product: 1, renew: 1\}}
& 32
& 125
& 76.6
& 68.8
& 46.4 \\

\bottomrule
\end{tabularx}

\caption{
Qualitative appendix example for the query  qid: "319" out-of-domain from  robust04 dataset
All query representations are shown after applying the same BM25 analyzer used at retrieval time.
Thus, terms appear in their analyzed form, e.g., \texttt{ongo}, \texttt{sourc}, \texttt{energi}, and \texttt{renew}.
The table reports the analyzed term-count dictionary, the number of unique analyzed terms, the total number of generated analyzed terms, and the resulting retrieval effectiveness and efficiency.
In this example, STORM produces a compact query specification and achieves the best nDCG@10 and Recall@1000, while remaining close to the original BM25 retrieval latency.
}
\label{tab:appendix_fuel_sources_full_case_study}
\end{table*}

\end{document}